\documentclass{article}
\usepackage{spconf,amsmath,graphicx, amsfonts}

\usepackage{multirow}
\usepackage{xcolor}
\usepackage{soul}
\usepackage{enumitem}
\usepackage[flushleft]{threeparttable}
\PassOptionsToPackage{hyphens}{url}\usepackage{hyperref}
\hypersetup{colorlinks=true}
\usepackage{changes}
\usepackage{todonotes}
\usepackage{booktabs}
\usepackage{tabularx}
\usepackage{float}
\usepackage{listings}
\newcommand\vect[1]{\mathbf{#1}}

\title{Textual Echo Cancellation}
\ifthenelse{\isundefined{\makeblind}}
{
\name{Shaojin Ding \quad Ye Jia \quad Ke Hu \quad Quan Wang}
\address{Google LLC, USA \\
{
\small
    \{
    \href{mailto:shaojinding@google.com}{\nolinkurl{shaojinding}},
    \href{mailto:jiaye@google.com}{\nolinkurl{jiaye}},
    \href{mailto:huk@google.com}{\nolinkurl{huk}},
    \href{mailto:quanw@google.com}{\nolinkurl{quanw}}
    \}
    {\tt @google.com}
}
}
}
{
\name{BLIND}
\address{BLIND}
}

\begin{document}
\ninept
\maketitle
\begin{abstract}
In this paper, we propose Textual Echo Cancellation (TEC) --- a framework for cancelling the text-to-speech (TTS) playback echo\footnote{TTS playback denotes the synthesized TTS voice, and TTS playback echo denotes the reverberated TTS voice captured by the microphone.} from overlapping speech recordings. Such a system can largely improve speech recognition performance and user experience for intelligent devices such as smart speakers, as the user can talk to the device while the device is still playing the TTS signal responding to the previous query. We implement this system by using a novel sequence-to-sequence model with multi-source attention that takes both the microphone mixture signal and source text of the TTS playback as inputs, and predicts the enhanced audio. Experiments show that the textual information of the TTS playback is critical to enhancement performance. Besides, the text sequence is much smaller in size compared with the raw acoustic signal of the TTS playback, and can be immediately transmitted to the device or ASR server even before the playback is synthesized. Therefore, our proposed approach effectively reduces Internet communication and latency compared with alternative approaches such as acoustic echo cancellation (AEC).

\vspace{0.2em}
\noindent\textbf{Code:} {\scriptsize \url{https://github.com/wq2012/tec}} \\
\textbf{Models:} {\scriptsize \url{https://huggingface.co/collections/wq2012/textual-echo-cancellation}} \\
\textbf{Demo:} {\scriptsize \url{https://huggingface.co/spaces/wq2012/tec}}
\end{abstract}
\section{Introduction}
\label{sec:intro}

\begin{figure}[t]
	\centering
	\includegraphics[width=0.4\textwidth]{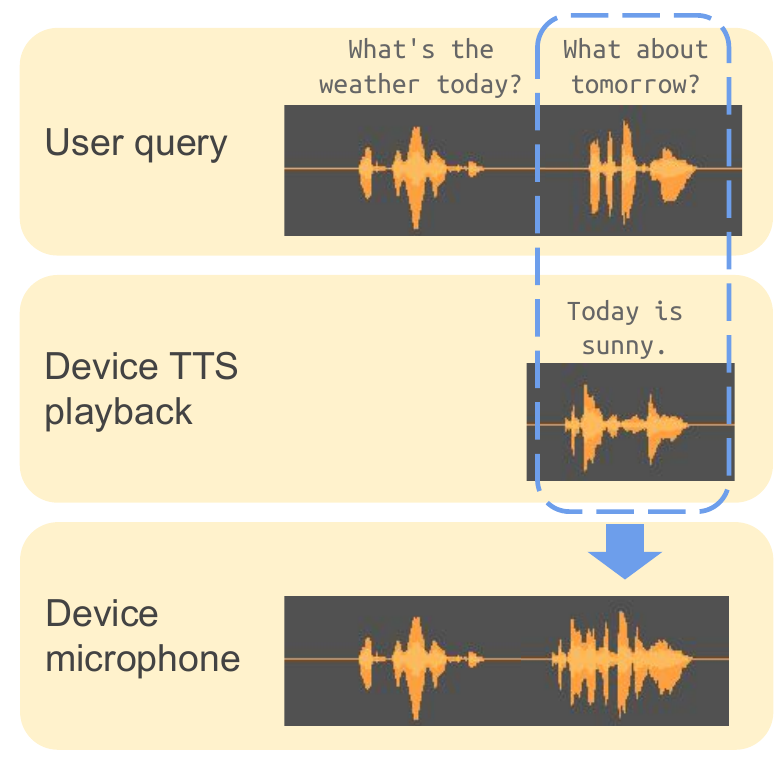}
	\vspace{-6pt}
	\caption{Acoustic echoes caused by TTS playback overlapping with user query.}
	\label{fig:use_case}
	\vspace{-14pt}
\end{figure}

Intelligent devices with speech interaction features have become popular in recent years, such as mobile devices and smart home speakers. In a typical user interaction, the user first issues a query to the device, then the device responds with synthesized speech; after hearing the response, the user may issue the next query. However, in some scenarios, the user may want to self-correct the previous query or impatiently issue a new query before the device finishes playing the synthesized response. When the user and the device talk at the same time, the acoustic echoes become a challenge for accurate speech recognition. For example, the user could first ask ``What's the weather today". While the smart speaker plays synthesized response ``Today is sunny", the user may interrupt impatiently with ``What about tomorrow", as illustrated in Fig.~\ref{fig:use_case}. In this case, it is very difficult for the device to correctly recognize the query ``What about tomorrow" due to the overlaps between query and TTS playback echo.

One of the most straightforward and well-developed approaches to solve this problem is acoustic echo cancellation (AEC)~\cite{hansler2005acoustic, benesty2001advances, zhang2018deep, fazel2019deep, lei2019deep, benesty2011perspective, enzner2014acoustic}. Conventional signal-processing based AEC approaches~\cite{hansler2005acoustic, benesty2001advances, benesty2011perspective, enzner2014acoustic} usually use adaptive filtering to estimate the echo path between the speaker and the microphone, and then the mixture signal from microphone is combined with the estimated echo path to produce the enhanced signal. More recently, model based AEC approaches~\cite{zhang2018deep, fazel2019deep, lei2019deep} have been shown to significantly boost the performance. These models take the microphone mixture signal along with the echo signal as the input and are trained to predict a mask, which is then applied to the microphone mixture signal to produce the enhanced signal.

However, for many practical applications, conventional AEC models are subject to a number of restrictions. First, under the intelligent device settings, there is no way to acquire the actual TTS playback echo since the room configurations are unknown and vary from user to user. As an approximation of the echo, we can directly use the TTS playback. However, there exists a mismatch between the actual echo and the TTS playback, which makes the AEC performance vary in real world conditions. Second, as speech synthesis is a computationally intensive task, TTS playback is usually streamed from a TTS server to the user device to achieve lower latency. However, most of existing AEC systems depend on entire TTS playback when producing the enhanced signal. If these systems run on the device, they cannot start running until the end of TTS streaming, which may introduce significantly high latency to ASR. If AEC and ASR are implemented on servers, then AEC would require the TTS service to stream the TTS playback to the AEC server as side input, which introduces additional Internet traffic.

\begin{figure*}
	\centering
	\includegraphics[width=0.8\textwidth]{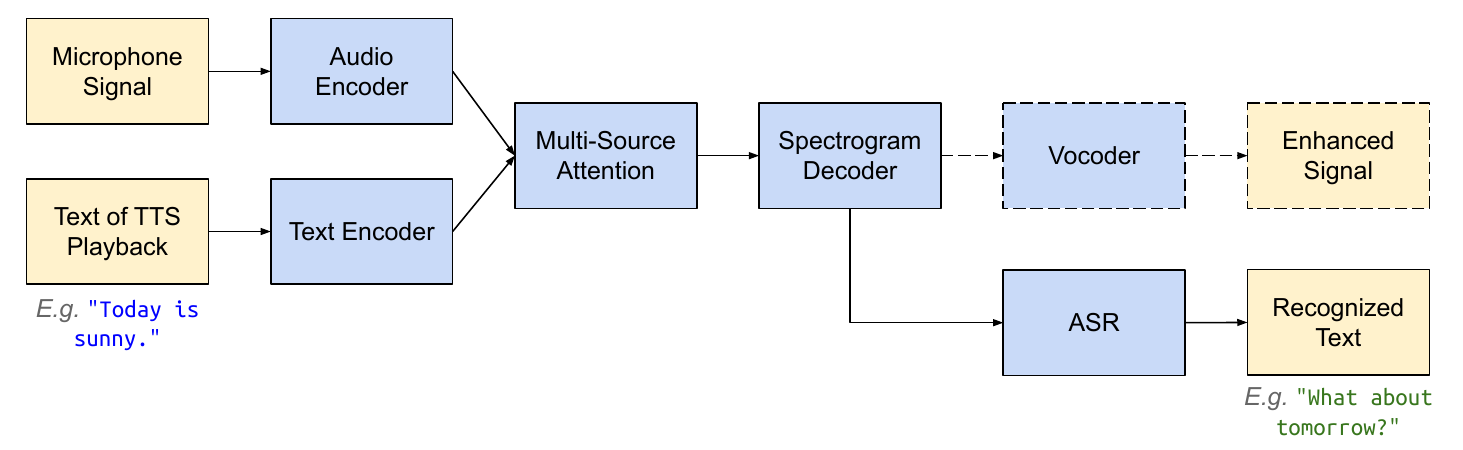}
	\vspace{-6pt}
	\caption{Diagram of the textual echo cancellation framework.}
	\label{fig:overall}
	\vspace{-16pt}
\end{figure*}


Other potential solutions to this problem are speech separation~\cite{hu2012unsupervised,huang2014deep,du2014speech,hershey2016deep,luo2018tasnet} and speaker extraction (\emph{a.k.a.}  voice filtering)~\cite{wang2018deep,delcroix2018single,wang2019voicefilter}. Speech separation models can directly separate two or more sources from the microphone mixture signal. However, these models require an output channel selection step after the separation to correctly keep the user's query by employing an additional speaker verification system. On the other hand, for example, the VoiceFilter system proposed in~\cite{wang2019voicefilter} separates the voice of a target speaker from multi-speaker signals, by making use of a reference signal from the target speaker. VoiceFilter assumes we have speech samples from the target user, such that we can build an embedding vector of the user's voice characteristics, and use this embedding vector as an auxiliary input to remove any signal that does not belong to the target user. However, if the user does not provide audio samples to enroll on the device, VoiceFilter is not feasible. Also, if the TTS playback voice is similar to the voice of the user, it is challenging for VoiceFilter to remove the TTS from the microphone mixture signal.

Previous studies also explored the use of text information in speech enhancement~\cite{kinoshita2015text, le2013text, erdogan2015phase}. However, our approach differs from these works in several aspects. \cite{kinoshita2015text} is the most relevant work to ours since they consider the use of text information in a speech enhancement system based on deep neural network. However, their model requires an extra algorithms~\cite{kinoshita2015text} to align speech signal and text, and these alignment algorithms are usually not very robust to background noise, as illustrated in~\cite{king2012noise, zhang2014one}. By contrast, our method achieves the alignment by the attention mechanism in the seq2seq model, which avoids the need of alignment algorithms. The other two studies share the idea of using text information in speech enhancement, but our approach is significantly different in either model (\cite{le2013text} uses non-negative matrix factorization) and how to incorporate text information (\cite{erdogan2015phase} uses a auxiliary speech recognizer).


To address these limitations, we propose a novel framework named as \textbf{Textual Echo Cancellation} (TEC)\ifthenelse{\isundefined{\makeblind}}{\footnote{Audio samples are available at \url{https://google.github.io/speaker-id/publications/TEC}}}{}. Instead of using the TTS playback or the user's speaker embedding as side input, we use the source text of the TTS playback. Comparing to the TTS playback, the source text is much smaller in size. Consequently, it can be efficiently transmitted between servers, or from server to device. This will avoid extra Internet communications and latency. The proposed approach is implemented using seq2seq modeling with multi-source attention~\cite{hu2020deliberation, xia2017deliberation, currey2018multi}, as shown in Fig.~\ref{fig:overall}. Our model takes two sequences as input:
The Mel-spectrogram of the mixture audio recorded from the microphone, and the source text corresponding to the TTS playback.
An audio encoder and a text encoder are used to extract representations for the two input sequences, respectively.
A multi-source attention attends to both the encoded speech and the encoded text, and outputs a fixed-dimensional context vector at each decoding step. Finally, the decoder consumes the context vectors and autoregressively predicts a Mel-spectrogram corresponding to the enhanced signal. The main contributions of our work are outlined as below:

\begin{itemize}
\vspace{-0.3em}
    \item We propose a novel seq2seq model for echo cancellation, which takes microphone mixture signals along with the source text of the TTS playback as the inputs, and produces enhanced audio signal. Compared to conventional echo cancellation approaches, our model avoids the need of TTS playback signal, reducing Internet communication and latency for smart speaker devices.
    \item We utilize a multi-source attention mechanism in our proposed seq2seq model. It generates a fix-length context vector during each decoding step based on the hidden representations extracted from microphone mixture signals and source text, thus incorporating information from multiple sources.
    \item We conduct extensive experiments on LibriTTS and VCTK datasets to evaluate the proposed model. We implement a conventional signal-processing AEC baseline and two seq2seq AEC baselines, and we compare the proposed model against them under two different conditions that mimic practical use cases of smart speaker devices.
\end{itemize}


The rest of the paper is organized as follows. In Section~\ref{sec:method}, we will give detailed description of different components of the proposed framework. In Section~\ref{sec:experiments}, we describe our experimental setup, including the data, model parameters, metrics, and results. Finally, we draw the conclusions and point out potential future work in Section~\ref{sec:conclusions}.

\section{Textual echo cancellation}
\label{sec:method}

\subsection{Overview}
A diagram illustrating the textual echo cancellation framework is shown in Fig~\ref{fig:overall}. Suppose we have a feature sequence $\vect{x}\in\mathbb{R}^{T_x\times D_m}$ for the microphone mixture signal, an embedding sequence $\vect{y}\in\mathbb{R}^{T_y\times D_e}$ representing the source text of the TTS playback, and a feature sequence $\vect{z}\in\mathbb{R}^{T_z\times D_m}$ for the user's actual clean speech. Here $T_x$, $T_y$, and $T_z$ are the lengths of the sequences $\vect{x}$, $\vect{y}$, $\vect{z}$, respectively, $D_m$ is the number of Mel-filterbanks (\emph{e.g.} 128 Mel-filterbanks in this work), and $D_e$ is the dimension of the text embedding (\emph{e.g.} 512-dimensional pre-trained phoneme embedding in this work). Our model consists of three modules: (1) an audio encoder, (2) a text encoder, and (3) an autoregressive decoder with a multi-source attention mechanism.

First, the audio encoder takes the microphone feature sequence as the input and produces a hidden representation:

\begin{equation}
    \vect{h_x} = \mathrm{Encoder_{audio}}(\vect{x})
\end{equation}

\noindent
Similarly, the text encoder takes the the source text sequence as the input and produces another hidden representation:

\begin{equation}
    \vect{h_y} = \mathrm{Encoder_{text}}(\vect{y})
\end{equation}

\noindent
Finally, the decoder autoregressively predicts the Mel-spectrogram of the enhanced signal using the attention context computed based on the two encoder outputs:

\begin{equation}
    \vect{\hat{z}}^{t} = \mathrm{Decoder}(\vect{\hat{z}}^{t-1}, \vect{h_x}, \vect{h_y})
\end{equation}

\noindent
The predicted Mel-spectrogram can be directly consumed by an ASR model or other downstream components such as vocoders. We will describe each module with details in the following subsections. The hyperparameters of each module are shown in Table~\ref{table:arch}.

\vspace{-4pt}
\subsection{Audio encoder}

The audio encoder converts a Mel-spectrogram sequence to a hidden representation sequence. Following \cite{biadsy2019parrotron},
we use an encoder composed of two convolutional layers, one bi-directional convolutional LSTM layer (Bi-CLSTM)~\cite{xingjian2015convolutional, schuster1997bidirectional}, and three bi-directional LSTM (Bi-LSTM) layers. A convolutional layer has 32 kernels, each of which has a shape of $3\times3$ in time $\times$ frequency and a stride of $2\times2$, followed by ReLU activations and batch normalization~\cite{ioffe2015batch}, capturing the local temporal context information. Meanwhile, a convolutional layer reduces the time resolution by a factor of 2, which also reduces the computational cost in the following layers. The Bi-CLSTM and Bi-LSTM layers extract high-level frequency-wise features and capture long-term temporal context information. Each of these layers have 256 units in each direction, followed by ReLU activations and batch normalization. Additionally, the Bi-CLSTM layer has $1\times3$ kernels with a stride of $1\times1$. As a result, the final output sequence of the audio encoder has a dimension of 512 and is four times shorter compared with the input sequence.

\vspace{-4pt}
\subsection{Text encoder}

The text encoder converts text sequences (represented by phonemes) to hidden representation sequences. Each of the input phoneme is first represented by a pre-trained 512-dimensional embedding. Then the embedding sequence is passed through three convolutional layers and one Bi-LSTM layer, following~\cite{shen2018natural}. Each convolutional layer has 512 kernels, and each kernel has a shape of $5\times1$ and stride of $1\times1$, followed by ReLU activations and batch normalization. Each kernel in the convolutional layers spans 5 phonemes, modeling the local context information (\emph{e.g.}, N-grams). The Bi-LSTM layer has 256 units in each direction, followed by ReLU activations and batch normalization, resulting in a 512-dimensional text encoder output sequence.

\vspace{-4pt}
\subsection{Decoder}

\begin{table}[t]
\begin{center}
\caption{Hyperparameters of the TEC network.}
\label{table:arch}
\resizebox{\columnwidth}{!}{
\begin{tabular}{c|c|c}
\hline
\multirow{2}{*}{Spectral analysis} & \multicolumn{2}{c}{frame length: 50 ms; frame shift: 12.5 ms;}  \\
& \multicolumn{2}{c}{128 Mel-filterbanks} \\
\hline
\multirow{5}{*}{Audio encoder} & \multirow{2}{*}{Conv layers $\times$ 2} & 32 3$\times$3 kernel with 2$\times$2 stride; \\
& & ReLU; batch norm\\
\cline{2-3}
& \multirow{2}{*}{Bi-CLSTM $\times$ 1} & 256 units per direction; \\
& & 1$\times$3 kernel with 1$\times$1 stride \\
\cline{2-3}
& Bi-LSTM $\times$ 3 & 256 units per direction \\
\hline
\multirow{4}{*}{Text encoder} & Text embedding & Pre-trained model; 512-dim \\
\cline{2-3}
& \multirow{2}{*}{Conv layers $\times$ 3} & 512 5$\times$1 kernel with 1$\times$1 stride; \\
& & ReLU; batch norm\\
\cline{2-3}
& Bi-LSTM $\times$ 1 & 256 units per direction \\
\hline
\multirow{2}{*}{Attention} & \multirow{2}{*}{Multi-source attention} & GMM attention for each source;  \\
 & & 128-dim attention context \\
\hline
\multirow{10}{*}{Decoder} & \multirow{2}{*}{PreNet} & fully-connected layer $\times$ 2 \\
 & & 256 neurons; ReLU \\
\cline{2-3}
 & LSTM $\times$ 2 & 256 units\\
\cline{2-3}
 & \multirow{2}{*}{Linear (Mel)} & fully-connect layer $\times$ 1 \\
 & & 128 neurons; no activation \\
\cline{2-3}
 & \multirow{2}{*}{Linear (stop token)} & fully-connect layer $\times$ 1 \\
 & & 2 neurons; no activation \\
\cline{2-3}
 & \multirow{3}{*}{PostNet} & Conv layers $\times$ 5 \\
 & & 512 5$\times$1 kernel with 1$\times$1 stride; \\
 & & TanH; batch norm \\
\hline
\end{tabular}}
\end{center}
\vspace{-20pt}
\end{table}

\begin{figure}
	\centering
	\includegraphics[width=0.4\textwidth]{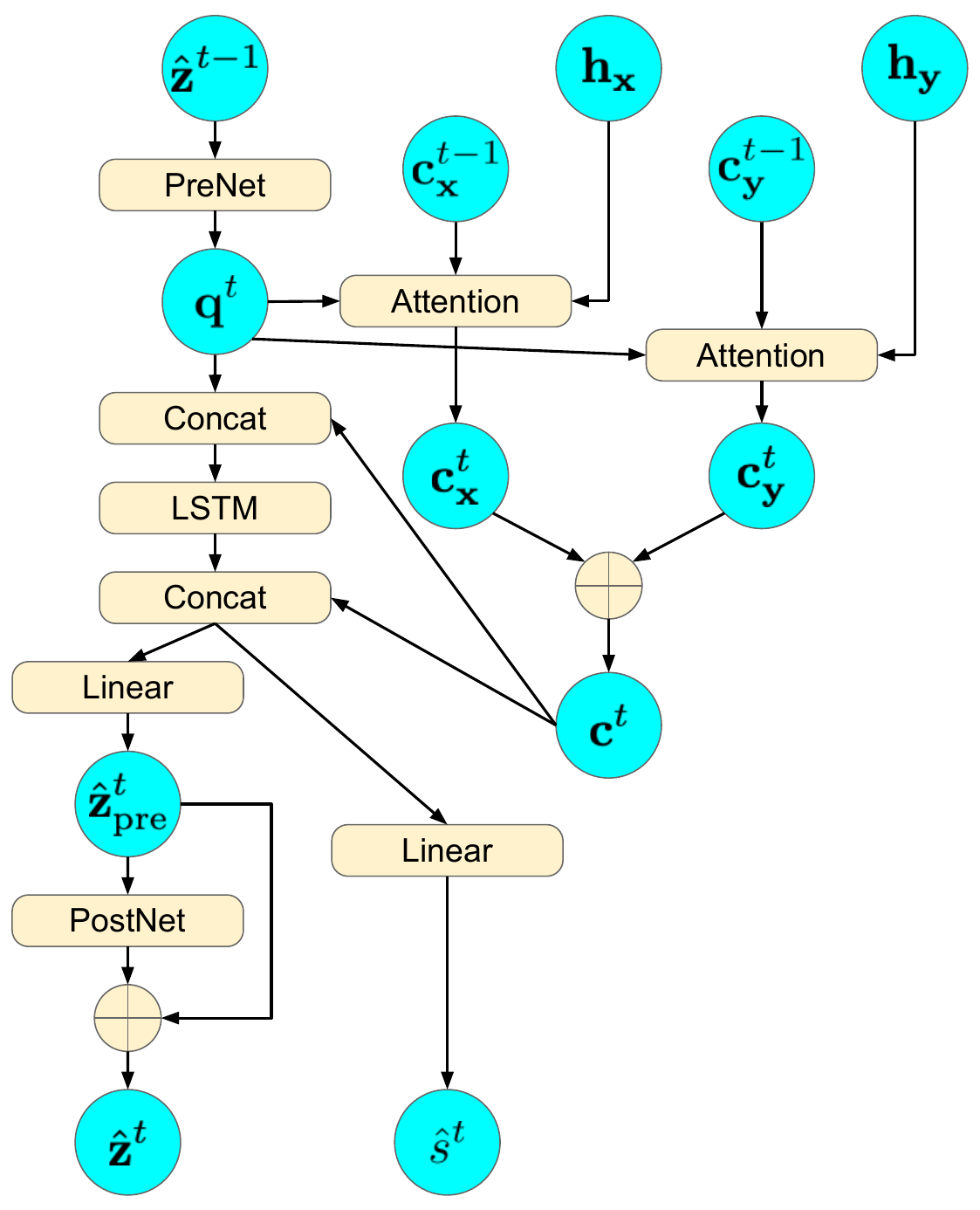}
	\caption{Diagram of the decoder with multi-source attention.}
	\label{fig:decoder}
	\vspace{-10pt}
\end{figure}

The decoder is an autoregressive recurrent neural network coupled with a multi-source attention mechanism (see Section~\ref{sec:msa}), as illustrated in Fig.~\ref{fig:decoder}. It takes the encoded sequences produced by the audio and text encoders as the inputs and generates the 128-dimensional enhanced Mel-spectrogram as a prediction of the user's clean speech signal.
We follow the same decoder architecture as in Tacotron~2~\cite{shen2018natural}. During each decoding step $t$, the prediction from the previous decoding step $\vect{\hat{z}}^{t-1}$ is fed to a pre-net containing two fully-connected layers of 256 neurons along with ReLU activations:
\begin{equation}
    \vect{q}^{t} = \mathrm{PreNet}(\vect{\hat{z}}^{t-1})
\end{equation}

\noindent
which is essential for learning attentions~\cite{shen2018natural}. Then the multi-source attention mechanism computes a 128-dimensional attention context vector $\vect{c}^t$ using the pre-net output, the attention context during the previous step, and the two encoded sequences from the two encoders:

\begin{equation}
    \vect{c}^t = \mathrm{MultiSourceAttention}(\vect{q}^t, \vect{c}^{t-1}, \vect{h_x}, \vect{h_y})
\end{equation}

\noindent
Next, the pre-net output and the attention context vector are concatenated and passed through two uni-directional LSTM layers with 256 units. The LSTM outputs are then concatenated again with the attention context vector and fed to a linear transformation of 128 units, resulting in a predicted Mel-spectrogram frame for the user's clean speech:

\begin{equation}
    \vect{\hat{z}}_\mathrm{pre}^{t} = \mathrm{Linear} \Big( \mathrm{LSTM}(\vect{q}^{t}, \vect{c}^{t}), \vect{c}^{t} \Big)
\end{equation}

\noindent
As the generated Mel-spectrogram from the seq2seq model is not a frame-synchronous estimate of $\vect{z}^t$, we also need the network to predict if the autoregressive generating process should stop at each decoding step, \emph{i.e.,} a 0/1 stop token $s^t$. Finally, to incorporate residuals in predicted Mel-spectrogram, these predictions are passed through 5-layer convolutional post-net, each layer having 512 kernels of shape $5\times1$ followed by batch normalization and tanh activation. The post-net predicts the residual that is added to the prediction, which has been shown to improve the Mel-spectrogram reconstruction~\cite{shen2018natural}:
\begin{equation}
    \vect{\hat{z}}^{t} = \mathrm{PostNet}(\vect{\hat{z}}_\mathrm{pre}^{t}) + \vect{\hat{z}}_\mathrm{pre}^{t}
\end{equation}

\vspace{-4pt}
\subsection{Multi-source attention mechanism}
\label{sec:msa}

We use a multi-source attention mechanism to summarize the encoded sequences from both audio and text encoders. The multi-source attention consists of two individual attentions for the two encoders, respectively, without sharing the weights between the two encoders. During each decoding step $t$, the two attentions first produce two fixed-length attention contexts:

\begin{equation}
    \vect{c}_{\vect{x}}^t = \mathrm{Attention_{audio}}(\vect{q}^t, \vect{c}_{\vect{x}}^{t-1}, \vect{h_x})
\end{equation}

\begin{equation}
    \vect{c}_{\vect{y}}^t = \mathrm{Attention_{text}}(\vect{q}^t, \vect{c}_{\vect{y}}^{t-1}, \vect{h_y})
\end{equation}

\noindent
Then we obtain the final context vector by summing up the two context vectors:

\begin{equation}
    \vect{c}^t = \vect{c}_{\vect{x}}^t + \vect{c}_{\vect{y}}^t
\end{equation}

\noindent
There are other strategies to combine the two context vectors, such as averaging, concatenation, and hierarchical attention combination~\cite{libovicky2017attention}. However, our preliminary results show that the difference between different combination strategies are minimal, so we use the simplest summation operation here. In addition, we use Gaussian mixture attention mechanism~\cite{graves2013generating} for both audio and text, which has been shown to achieve superior performance than conventional additive attention mechanism~\cite{bahdanau2014neural} on speech synthesis~\cite{he2019robust, skerry2018towards, polyak2019attention}.

\subsection{Model training and inference}

Following \cite{jia2018transfer}, during training, the model is optimized by minimizing the sum of the L1 and L2 distances computed from the output before and after the post-net. We apply the teacher-forcing training procedure (feeding in the correct output instead of the predicted output on the decoder side). As a result, we need to jointly minimize an extra cross-entropy loss to learn the stop token for model inference. The overall loss function of the proposed model is:


\begin{equation}
\begin{aligned}
    L = & ||\vect{\hat{z}_\mathrm{pre}} - \vect{z}||^2_2 + ||\vect{\hat{z}} - \vect{z}||^2_2 + \\
    & ||\vect{\hat{z}_\mathrm{pre}} - \vect{z}||_1 + ||\vect{\hat{z}} - \vect{z}||_1 + \\
    & \mathrm{CrossEntropy}(\vect{\hat{s}}, \vect{s})
\end{aligned}
\end{equation}

\noindent
where $\vect{\hat{s}}$ is the sequence of the predicted stop token and $\vect{s}$ is the sequence of the target stop token.

Once we have a trained model, we can pass the Mel-spectrogram of the microphone mixture signal along with the source text of the TTS playback to the model to acquire the Mel-spectrogram of the enhanced signal. The Mel-spectrogram can be directly consumed by downstream ASR. Additionally, we can also use a vocoder (\emph{e.g.} WaveNet~\cite{vanwavenet} or WaveRNN~\cite{kalchbrenner2018efficient}) to synthesize the waveform of the enhanced audio if it is needed for other downstream modules.

\section{Experiments}
\label{sec:experiments}
We conduct experiments under two different conditions to evaluate the proposed approach. In the first experiment, we consider the TTS voice being generated from a canonical speaker (denoted as \emph{single interfering voice condition}). Following this, we extend the TTS voice to be have multiple different speakers' identities in the second experiment (denoted as \emph{multiple interfering voices condition}), which is closer to real-world scenarios (\emph{e.g.} personalized playback voice in smart speaker devices) but more challenging.

\subsection{Datasets}

\begin{table*}[ht]
\begin{center}
\caption{Data configuration for our experiments. This table shows how the microphone signal mixtures were generated. For example, in single interfering voice condition, the synthetic training set was mixed using LibriTTS training set and LJ Speech training set.}
\label{table:synthetic}
\resizebox{0.65\textwidth}{!}{
\begin{tabular}{c|c|c|c}
\hline
Condition & Synthetic subset & User's speech & Interfering speech \\
\hline
\multirow{3}{*}{Single interfering voice} & training & LibriTTS training & LJ Speech training \\
 & test-clean & LibriTTS test-clean & LJ Speech test \\
 & test-other & LibriTTS test-other & LJ Speech test \\
\hline
\multirow{3}{*}{Multiple interfering voices} & training & LibriTTS training & VCTK training \\
 & test-clean & LibriTTS test-clean & VCTK test \\
 & test-other & LibriTTS test-other & VCTK test \\
\hline
\end{tabular}}
\end{center}
\vspace{-16pt}
\end{table*}

Following previous echo cancellation studies~\cite{zhang2018deep, lei2019deep, fazel2019deep, fazel2020cad}, we use synthetic data for the evaluations. In both conditions, we use the LibriTTS dataset~\cite{zen2019libritts} for the user query. To produce microphone mixture signal, we mix utterances from LibriTTS with the utterances from the LJ Speech dataset~\cite{ljspeech17} and the CSTR VCTK dataset~\cite{veaux2016superseded} in the single/multiple interfering voice(s) conditions, respectively (see Section~\ref{sec:synthetic}). Comparing against TIMIT dataset that is commonly used in previous studies~\cite{zhang2018deep, lei2019deep, fazel2019deep, fazel2020cad}, these datasets contain continuous sentences instead of just the recording of ten digits, which is more appropriate in simulating the practical use cases.

The LibriTTS dataset consists of 585 hours of audio book speech data from 2,456 speakers. The dataset is divided into three parts: 555 hours of training sets, 15 hours of development sets, and 15 hours of testing set. Each of them contains both clean and noisy speech.

The LJ Speech dataset has 24 hours clean audio book speech data from a single speaker. The original LJ Speech dataset does not have training and testing subsets. For evaluation purpose, we randomly selected 90\% of the utterances as the training set and the remaining 10\% as the testing set.

The CSTR VCTK dataset contains 44 hours of clean speech from 109 speakers. Similarly, we randomly selected 90\% of the utterances from each speaker as the training set and the remaining 10\% as the testing set, since there are no official training and testing subsets.

\subsection{Generating synthetic microphone mixture signal}
\label{sec:synthetic}

We use the acoustic signal model described in~\cite{fazel2020cad, fazel2019deep} to generate synthetic microphone mixture signals. In their model, the microphone mixture signal $x(n)$ is generated as:

\begin{equation}
    \label{eq:rir}
    x(n) = z(n) + y(n)*h(n)
\end{equation}

\noindent
where $z(n)$ is user's speech signal, $y(n)$ is TTS playback, $h(n)$ is the room impulse response (RIR), and $*$ is the convolutional operation. A total number of 3 million RIR were generated using a room simulator~\cite{lippmann1987multi, ko2017study, kim2017generation} to cover different reverberation conditions. With Eq.~\ref{eq:rir}, we generated three subsets of synthetic data: (1) training, (2) test-clean, and (3) test-other, as shown in Table~\ref{table:synthetic}.

For each user's speech utterance, we randomly chose an interfering utterance and followed the above model to generate a microphone mixture signal with a 0-dB Signal-Noise-Ratio (SNR). As a result, the number of mixtures in the synthetic datasets is the same as the number of utterances in the LibriTTS dataset. Additionally, we padded the two utterances to have the same length to handle the duration difference between two utterances.

For interfering speech, we use the same TTS speakers during model training and evaluation. This is based on the fact that for smart home speaker devices, there is typically a fixed set of TTS voice options. However, for user's speech, we use different sets of speakers during training and evaluation (\emph{e.g.}, LibriTTS train vs. dev/test sets), since the user of the device is unknown at runtime.

\vspace{-4pt}
\subsection{Metrics}
To evaluate the proposed approach, we consider three metrics: (1) Word Error Rate (WER), (2) Mel-Cepstral Distortion (MCD)~\cite{kubichek1993mel}, and (3) Mean Opinion Score (MOS) of the naturalness of the enhanced speech audio. Additionally, we also estimate the side input size and computational complexity for different approaches.

\vspace{-6pt}
\subsubsection{Word Error Rate}
\vspace{-4pt}
As described in Section~\ref{sec:intro}, the main purpose of our proposed approach is to improve the speech recognition performance of smart speaker devices when the user's query and the TTS playback echo have overlaps. As a result, we use WER as the major evaluation metric for our experiments. The speech recognizer we used for WER evaluation is a state-of-the-art model proposed in~\cite{park2020improved}, which is trained on the LibriSpeech~\cite{panayotov2015librispeech} training set. We did not re-train the recognizer on the generated features.

\vspace{-6pt}
\subsubsection{Mel-Cepstral Distortion}
\vspace{-4pt}
MCD (dB) is a commonly used objective metric to evaluate the quality of the synthesized speech, which is defined as

\begin{equation}
    \textrm{MCD} = \frac{10}{\ln 10} \sum_{t=1}^{T_z} \sqrt{2\sum_{d=1}^{13} (\hat{z}_{t,d} - z_{t,d})^2}
\label{eq:mcd}
\end{equation}

\noindent
where $\hat{z}_{t,d}$ and $z_{t,d}$ are the $d$-th Mel-Frequency Cepstral Coefficient (MFCC) of the enhanced speech and the time-aligned\footnote{We use dynamic time warping~\cite{salvador2007toward} for time alignment.} target speech at the $t$-th time step, respectively. In this paper, we used 13 MFCCs (skipping MFCC$_0$, which is energy) to compute MCD. Lower MCD indicates that the enhanced speech is more similar to the target clean speech.

We did not include the metrics that are used in conventional echo cancellation approaches (\emph{e.g.} echo return loss enhancement metric~\cite{enzner2014acoustic}, perceptual evaluation of speech quality~\cite{rix2001perceptual}) for two reasons. First, the downstream ASR model can directly take the Mel-spectrogram as the input, and therefore, generating waveform becomes redundant and may cause extra distortions. Second, the waveform of the enhanced signal is generated using generative neural vocoder models, and the generated waveform can be very different from the target waveform, even if the linguistic content of the waveforms are exactly the same.. As a result, these waveform based metrics become ill-defined in our case.

\vspace{-6pt}
\subsubsection{Speech naturalness Mean Opinion Score}
\vspace{-4pt}
We measured the speech naturalness of the enhanced signal with a 5-point Mean Opinion Score (1-bad; 5-excellent). For each system, we randomly chose 1,000 utterances from test-clean/test-other subsets for MOS evaluation. All the utterances were synthesized using a separately trained WaveRNN model~\cite{kalchbrenner2018efficient}. Each sample was rated by six raters, and each evaluation was conducted independently: the outputs of different models were not compared directly.

\vspace{-6pt}
\subsubsection{Side input size and computational complexity}
\vspace{-4pt}
We also computed the side input size and the computational complexity to evaluate the resources that are required for practical applications. To measure the computational complexity, we estimate the number of floating-point operations (FLOPS) required following~\cite{hu2020deliberation}:

\begin{equation}
    \mathrm{FLOPS} = M_\mathrm{audio} \cdot T_x + M_\mathrm{text} \cdot T_y + M_\mathrm{dec} \cdot T_z + \mathrm{FLOPS_{atten}}
\end{equation}

\noindent
where $M_\mathrm{audio}$, $M_\mathrm{text}$, and $M_\mathrm{dec}$ are number of model parameters of the audio encoder, text encoder, and decoder, respectively. $T_x$, $T_y$, and $T_z$ are the sequence lengths of the Mel-spectrogram of the microphone mixture signal, text embedding, and user’s actual clean speech signal, respectively. $\mathrm{FLOPS_{atten}}$ is the FLOPS required for the multi-source attention layer. For each attention source, we first multiply the size of the attention source matrix with $T_x$/$T_y$, and multiply the size of query matrix with $T_z$, and then the FLOPS for multi-source attention is computed as the sum of the two.

\begin{table*}[ht]
\begin{center}
\caption{Word Error Rate (WER), Mel-Cepstral Distortion (MCD), and speech naturalness Mean Opinion Score (MOS) evaluation results of the single speaker interfering voice and multiple interfering voices conditions. The MOS is presented with 95\% confidence intervals. We also include the size of the side input in kilobytes (the TTS playback echo or the TTS source text) and the floating point operations per second in Giga (GFLOPS).}
\label{table:results}
\resizebox{0.95\textwidth}{!}{
\begin{threeparttable}
\begin{tabular}{c|c|cc|cc|cc| c | c}
\hline
\multirow{3}{*}{Condition} & \multirow{3}{*}{Method} & \multicolumn{2}{c|}{WER (\%)} & \multicolumn{2}{c|}{MCD} & \multicolumn{2}{c|}{MOS} & \multirow{3}{*}{Side input (KB)} & \multirow{3}{*}{GFLOPS} \\
\cline{3-8} & & test- & test- & test- & test- & \multirow{2}{*}{test-clean} & \multirow{2}{*}{test-other} & &  \\
 & & clean & other & clean & other &  &  & & \\
\hline
Ground-truth LibriTTS & - & 2.30 & 4.50 & 0.00 & 0.00 & 4.43 $\pm$ 0.04 & 3.82 $\pm$ 0.06 & - & - \\
\hline
\multirow{4}{*}{Single interfering voice} & Microphone signal & 89.9 & 120.5  & 18.83 & 21.44 & - & - & - & - \\
\cline{2-10}
 & AEC-NLMS & 48.6 & 60.1 & 12.26 & 12.57 & 1.95 $\pm$ 0.10 & 1.28 $\pm$ 0.09 & 310 & 0 \\
 & Vanilla-Seq2seq & 25.4 & 54.0 & 7.85 & 8.84 & 1.99 $\pm$ 0.06 & 1.47 $\pm$ 0.05 & 0 & 6.32\\
 & AEC-Seq2seq & 8.30 & 24.3 & 6.38 & 7.07 & 2.77 $\pm$ 0.07 & 1.90 $\pm$ 0.06 & 310 & 9.51 \\
 & TEC (proposed) & 15.5 & 39.8 & 7.51 & 8.54 & 2.20 $\pm$ 0.07 & 1.65 $\pm$ 0.06 & 0.10 & 7.27 \\
\hline
\multirow{4}{*}{Multiple interfering voices} & Microphone signal & 29.7 & 44.6  & 10.75 & 12.88 & - & - & - & - \\
\cline{2-10}
 & AEC-NLMS & 15.5 & 35.5 & 6.57 & 8.13 & 2.06 $\pm$ 0.11 & 1.60 $\pm$ 0.08 & 230 & 0 \\
 & Vanilla-Seq2seq & 19.7 & 38.7 & 7.53 & 8.87  & 2.16 $\pm$ 0.07 & 1.50 $\pm$ 0.05 & 0 & 6.32 \\
 & AEC-Seq2seq & 6.90 & 19.8 & 5.04 & 5.72 & 2.90 $\pm$ 0.07 & 2.03 $\pm$ 0.07 & 230 & 8.62 \\
 & TEC (proposed) & 14.8 & 32.5 & 6.46 & 7.71 & 2.39 $\pm$ 0.07 & 1.70 $\pm$ 0.06 & 0.06 & 6.90 \\
\hline
\end{tabular}
\item[$\star$] The side input size and GFLOPS in the two conditions are different since the average lengths of the echo signal are different in the two conditions.
\end{threeparttable}}
\end{center}
\vspace{-20pt}
\end{table*}

\vspace{-4pt}
\subsection{Implementation details}

We implemented the model using the Lingvo~\cite{shen2019lingvo} framework in TensorFlow~\cite{abadi2016tensorflow}. Our model was trained on 2$\times$2 Tensor
Processing Units (TPU) slices with a global batch size of 32. During training, we use Adam optimizer~\cite{kingma2014adam} with $\beta_1=0.9$, $\beta_2=0.999$, and $\epsilon=10^{-6}$. We set the initial learning rate to $10^{-4}$ and exponentially decays to $10^{-5}$ after 50,000 iterations.

\subsection{Results}

In each condition, we compared the proposed approach against three baselines that we implemented:
(1) AEC-NLMS: A conventional signal-processing AEC algorithm based on normalized least mean square (NLMS) algorithm~\cite{enzner2014acoustic} that is widely used in prior studies~\cite{zhang2018deep, fazel2020cad, fazel2019deep}. (2) Vanilla-Seq2seq: A sequence-to-sequence network that directly transforms the microphone mixture signal to enhanced signal without any side input using single attention. This network is similar to \cite{biadsy2019parrotron, jia2019direct} except for not using auxiliary decoders, and we use a Gaussian mixture attention for it. (3) AEC-Seq2seq: An end-to-end model with similar architecture as the proposed TEC model. We replaced the text encoder in the proposed approach with an audio encoder that takes the TTS playback as the input, which operates similarly to other model-based AEC models.


The WER, MCD, and MOS evaluation results of the two conditions are shown in Table~\ref{table:results}. We include the three measurements of the ground-truth LibriTTS test-set, acting as a performance upper bound. Under single interfering voice condition, we observed that TEC achieves 15.5\% WER, 7.51 MCD, and 2.20 MOS on test-clean subset as well as 39.8\% WER, 8.54 MCD, and 1.65 MOS on test-other subset. Under multiple interfering voices condition, TEC achieves 14.8\% WER, 6.46 MCD, and 2.39 MOS on test-clean subset as well as 32.5\% WER, 7.71 MCD, and 1.70 MOS on test-other subset. Comparing against the baseline systems, the WER, MCD, and MOS results in both conditions consistently suggest that our proposed approach significantly outperforms AEC-NLMS, corresponding to the observations obtained from prior studies that deep learning models usually achieve superior performance than signal-processing based AEC algorithms. In addition, TEC also achieves essential improvement than Vanilla-Seq2seq, indicating that the information of TTS playback is key in achieving reasonable echo cancellation performance. However, we found that TEC is not as good as AEC-Seq2seq. This is expected since TEC only uses the source text of the TTS playback instead of the TTS playback echo, and the source text contains less information about how the TTS playback actually sounds like. Although there is a performance gap between TEC and AEC-Seq2seq, the size of the side inputs and the GFLOPS of TEC are much lower than AEC-Seq2seq, which supports our argument that TEC is more efficient in terms of Internet communications and latency.

The performances of all the systems on test-clean subset are better than those on test-other subset, since the utterances in test-other subset have considerable background noises, which degrades the quality of the output enhanced audio. Additionally, it is interesting to observe that the WER, MCD and MOS of all the systems under multiple interfering voices condition are better than those under single interfering voice condition. A possible explanation of this observation is that the utterances in VCTK are much shorter than those in LJ Speech ($\sim$2 seconds vs. $\sim$7 seconds, in terms of average duration per utterance), and therefore, the interfered intervals under multiple interfering voices condition are much shorter than those under single interfering voice condition. It makes the test set under multiple interfering voices condition an easier case, which is also supported by our results that the WER and MCD of the microphone signal under this condition are lower than those under single interfering voice condition. Besides, most of the utterances in VCTK have a British English accent, while LibriTTS and LJ Speech  are dominated by American English accent. Both factors make it easier to separate the user's speech from the interfering speech under multiple interfering voices condition than that under single interfering voice condition.

\vspace{-6pt}
\section{Conclusions and future work}
\vspace{-4pt}
\label{sec:conclusions}
In this paper, we proposed textual echo cancellation, a framework to cancel the TTS playback echo from overlapped speech, which is useful when a user talks to an intelligent device while the device is still playing synthesized response to a previous query. Our proposed approach uses the source text as the side input instead of the TTS playback, which can be efficiently transmitted between servers and from server to device, thus largely reducing Internet communications and latency compared with conventional AEC-Seq2seq approaches. We conducted experiments under a single interfering voice condition and a multiple interfering voices condition. Our experimental results show that TEC significantly outperforms the baseline of not using any side input, indicating that the textual information of the TTS playback is critical to the enhancement performance. In addition, the side input size and GFLOPS of TEC are much lower than model based AEC-Seq2seq methods.

In our experiments, the performance of TEC is still not as good as that of AEC-Seq2seq. Several directions can be explored in the future. First, a second decoder for phonetic recognition can be added during training, which has shown to be effective in~\cite{biadsy2019parrotron, jia2019direct} for speech-to-speech conversion models. Additionally, alternative architectures such as frame-to-frame TEC models can be implemented and compared with our current multi-source attention sequence-to-sequence TEC model. Furthermore, in applications where there is no restrictions on computations and data transmissions (\emph{e.g.}, offline echo cancellation), we can consider both TTS playback and source text as the side inputs to the model, which may provide extra performance gains. Last but not least, we can train and evaluate the proposed approach on data in the wild instead of synthetic data. As mentioned in Section~\ref{sec:intro}, AEC-Seq2seq model is subjected to the mismatch problem between the TTS playback echo and the TTS playback. By contrast, TEC does not have such a problem, and therefore, we believe the gap between AEC-Seq2seq and TEC will become smaller on wild data (see Appendix~\ref{sec:open_source_codebase}--\ref{sec:open_source_results} for our open-source release).

\bibliographystyle{IEEEbib}
\clearpage
\bibliography{refs}

\clearpage
\onecolumn
\appendix
\sloppy
\raggedbottom

\section{Open-Source Codebase and Python Package}
\label{sec:open_source_codebase}

To make the Textual Echo Cancellation (TEC) framework transparent, accessible, and reproducible outside of proprietary infrastructure, we have developed and released a standalone open-source Python implementation. The source code, command-line pipelines, unit test suite, and documentation are publicly available on GitHub, and the library is distributed as an installable package on the Python Package Index (PyPI):

\vspace{0.3em}
\noindent
\begin{tabularx}{\textwidth}{l X}
\textbf{GitHub Repository:} & \url{https://github.com/wq2012/tec} \\
\textbf{PyPI Package:} & \url{https://pypi.org/project/textual-echo-cancellation/} (\texttt{pip3 install textual-echo-cancellation==0.2.3}) \\
\textbf{Pretrained Models:} & \url{https://huggingface.co/collections/wq2012/textual-echo-cancellation} (6 model repositories; see Appendix~\ref{sec:open_source_models}) \\
\textbf{Interactive Demo (Space):} & \url{https://huggingface.co/spaces/wq2012/tec} (see Section~\ref{sec:hf_space_demo}) \\
\textbf{Audio Demo Page:} & \url{https://google.github.io/speaker-id/publications/TEC/} \\
\textbf{License:} & Apache License 2.0
\end{tabularx}
\vspace{0.3em}

\subsection{Comparison Between Internal and Open-Source Implementations}
\label{sec:arch_comparison}

The experiments in the main body of this paper (Section~\ref{sec:experiments} and Table~\ref{table:results}) were originally conducted using Google's internal software stack, which relied on internal speech frontends, a pre-trained proprietary TTS phoneme embedding table, a 3-million-room internal acoustic simulator~\cite{kim2017generation}, a separately trained internal WaveRNN neural vocoder~\cite{kalchbrenner2018efficient}, and an internal LibriSpeech Noisy Student ASR model~\cite{park2020improved}. Because those internal dependencies cannot be distributed publicly, the open-source \texttt{textual-echo-cancellation} library re-implements the complete TEC modeling, data preparation, waveform synthesis, TFLite export, and evaluation stack from scratch using open-source \texttt{lingvo}~\cite{shen2019lingvo} and \texttt{tensorflow}~\cite{abadi2016tensorflow}. Table~\ref{table:internal_vs_opensource} summarizes the exact architectural and infrastructure correspondence between the original internal pipeline and the open-source reproduction.

\begin{table}[!htbp]
\caption{Detailed comparison between the original internal implementation (Section~\ref{sec:method}--\ref{sec:experiments}) and the standalone open-source reproduction (\texttt{textual-echo-cancellation}).}
\label{table:internal_vs_opensource}
\vspace{0.15cm}
\centering
\scriptsize
\begin{tabularx}{\textwidth}{l >{\raggedright\arraybackslash}X >{\raggedright\arraybackslash}X}
\toprule
\textbf{Component} & \textbf{Original Internal Implementation (Main Paper)} & \textbf{Open-Source Reproduction (\texttt{wq2012/tec})} \\
\midrule
\textbf{Modeling Framework} & Internal Google Lingvo / TensorFlow trained on $2\times 2$ Cloud TPU slices (global batch size 32, 50,000 steps). & Standalone open-source \texttt{lingvo} ($\ge 0.12.7$) and \texttt{tensorflow} ($\ge 2.9.3$) supporting CPU, GPU, and TFLite execution. \\
\midrule
\textbf{Acoustic Frontend} & 24~kHz sample rate, 50~ms frame length (1,200 samples), 12.5~ms frame shift (300 samples), 128 log-Mel filterbanks. & \texttt{WaveformProcessor} (\texttt{tec/waveform\_processor.py}): 24~kHz, 50~ms Hann window, 12.5~ms hop, 2,048-point FFT, 128 log-Mel bins (20--12,000~Hz, natural-log compression with magnitude floor $10^{-3}$). \\
\midrule
\textbf{Audio Encoder} & 2 strided $3\times 3$ Conv2D layers ($2\times 2$ stride, 32 channels), 1 Bi-CLSTM ($1\times 3$ kernel, 256 units/dir), 3 Bi-LSTM layers (256 units/dir). & \texttt{SpeechEncoderV1} (\texttt{tec/encoder\_speech.py}): Subclasses \texttt{lingvo.tasks.asr.encoder.AsrEncoder} with identical $3\times 3$ Conv2D pyramid ($4\times$ time reduction), 1 Bi-CLSTM (\texttt{num\_conv\_lstm\_layers=1}, $1\times 3$ kernel), and 3 Bi-LSTM layers (512-dim output). \\
\midrule
\textbf{Text Tokenizer \& Encoder} & Internal TTS text normalization and pre-trained 512-dim phoneme embedding table, followed by 3 Conv1D layers ($5\times 1$, 512 channels, BatchNorm, ReLU) and 1 Bi-LSTM (256 units/dir). & \texttt{CharTokenizer} (\texttt{tec/tokenizer.py}) + \texttt{TtsEncoderV2} (\texttt{tec/encoder\_text.py}): Self-contained 98-symbol printable ASCII character vocabulary (\texttt{VOCAB\_SIZE = 98}: \texttt{<pad>=0}, \texttt{</s>=1}, \texttt{<unk>=2} + all 95 printable ASCII characters \texttt{0x20}--\texttt{0x7E}) with learned 512-dim embedding, 3 Conv1D layers ($5\times 1$, 512 channels, BatchNorm, ReLU, dropout 0.5), and 1 Bi-LSTM (256 units/dir, zoneout 0.1). \\
\midrule
\textbf{Multi-Source Attention} & Independent 5-mixture Gaussian Mixture Model (GMM) monotonic attention~\cite{graves2013generating} per encoder stream with context summation $\vect{c}^t = \vect{c}^t_{\vect{x}} + \vect{c}^t_{\vect{y}}$. & \texttt{MultiSourceAttentionStep} (\texttt{tec/multi\_source\_attention\_steps.py}): Wraps \texttt{GmmMonotonicAttention} (5 mixtures, 128-dim hidden state) or \texttt{AdditiveAttention} for each source stream (\texttt{source\_0}, \texttt{source\_1}) and sums context vectors. \\
\midrule
\textbf{Spectrogram Decoder} & Tacotron-2~\cite{shen2018natural} Pre-Net ($2\times 256$ ReLU, dropout 0.5), 2-layer LSTM (256 units), linear Mel and stop-token heads, and 5-layer Conv1D Post-Net ($5\times 1$, 512 channels, BatchNorm, $\tanh$). & \texttt{MultiSourceFbeDecoderV1} and \texttt{FbeDecoderV1} (\texttt{tec/decoder.py}) + \texttt{PostEditConvNet} (\texttt{tec/layers.py}): Identical layer dimensions with frame reduction factor $r=4$, masked $L_1 + L_2 + \mathrm{BCE}$ loss, and step-aligned spectral residual skip connections for fast convergence. \\
\midrule
\textbf{Waveform Reconstruction} & Separately trained autoregressive WaveRNN vocoder~\cite{kalchbrenner2018efficient}. & Training-free phase-preserving Mel-band gain filtering (using microphone mixture STFT phase) and 100-iteration Griffin-Lim phase reconstruction in \texttt{WaveformProcessor.SpectrogramsToWaveforms}. \\
\midrule
\textbf{Room Simulator (RIR)} & 3 million room impulse responses from an internal room simulator~\cite{lippmann1987multi, ko2017study, kim2017generation}. & Analytic exponentially decaying room impulse response generator (\texttt{generate\_synthetic\_rir} in \texttt{tec/data\_prep.py}, $\mathrm{RT}_{60}=0.25\text{ s}$, 5~ms direct-path delay, \texttt{tail\_scale=0.05} yielding $+6.64\text{ dB}$ direct-to-reverberant ratio, FFT convolution via \texttt{scipy.signal.fftconvolve}). \\
\midrule
\textbf{ASR Engine for WER} & Internal LibriSpeech Noisy Student ASR model~\cite{park2020improved} consuming Mel-spectrograms. & Open-source \texttt{Qwen3-ASR-0.6B} (\texttt{Qwen3-ASR-0.6B-F16.gguf}) executed locally on CPU via \texttt{audio.cpp} on reconstructed 24~kHz WAV files. \\
\midrule
\textbf{On-Device Export} & Not included in original paper. & \texttt{tec/tflite\_export.py}: Exports any trained Lingvo model to a TensorFlow Lite (\texttt{.tflite}) FlatBuffer with dynamic-range quantization (enabled by default) and Flex (\texttt{SELECT\_TF\_OPS}) support. \\
\bottomrule
\end{tabularx}
\end{table}

\subsection{Package Structure and Module Reference}
\label{sec:module_reference}

The repository is organized into a core Python package (\texttt{tec/}), six command-line execution scripts (\texttt{scripts/}), and a comprehensive unit test suite (\texttt{tests/}) achieving 100\% test pass rate across 37 unit tests in 9 test files. Table~\ref{table:codebase_modules} documents each module in the open-source codebase and its correspondence to the sections and equations of this paper.

\begin{table}[H]
\caption{Module-by-module reference for the open-source \texttt{textual-echo-cancellation} package (\url{https://github.com/wq2012/tec}).}
\label{table:codebase_modules}
\vspace{0.15cm}
\centering
\scriptsize
\begin{tabularx}{\textwidth}{l >{\raggedright\arraybackslash}p{4.8cm} >{\raggedright\arraybackslash}X}
\toprule
\textbf{File Path} & \textbf{Key Classes / Functions} & \textbf{Paper Reference \& Functionality} \\
\midrule
\texttt{tec/waveform\_processor.py} & \texttt{WaveformProcessor}, \texttt{compute\_fft\_size} & Section~\ref{sec:method}.1 \& Table~\ref{table:arch}: 24~kHz STFT, 128-bin log-Mel filterbank extraction (\texttt{WaveformsToSpectrograms}), and Griffin-Lim / phase-preserving waveform synthesis (\texttt{SpectrogramsToWaveforms}). \\
\texttt{tec/tokenizer.py} & \texttt{CharTokenizer} & Section~\ref{sec:method}.3: Maps TTS source text strings to 98-symbol ASCII character token IDs and padding masks (\texttt{text\_to\_ids}, \texttt{ids\_to\_text}, \texttt{batch\_encode}). \\
\texttt{tec/encoder\_speech.py} & \texttt{SpeechEncoderV1} & Section~\ref{sec:method}.2, Eq.~(1) \& Table~\ref{table:arch}: Encodes 128-bin log-Mel spectrograms $\vect{x} \in \mathbb{R}^{T_x \times 128}$ into 512-dim hidden sequences $\vect{h_x}$ at $4\times$ reduced frame rate via 2 Conv2D layers, 1 Bi-CLSTM, and 3 Bi-LSTM layers. \\
\texttt{tec/encoder\_text.py} & \texttt{TtsEncoderV2} & Section~\ref{sec:method}.3, Eq.~(2) \& Table~\ref{table:arch}: Encodes TTS character IDs into 512-dim hidden representations $\vect{h_y}$ via 512-dim embeddings, 3 Conv1D layers ($5\times 1$), and 1 Bi-LSTM. \\
\texttt{tec/multi\_source\_attention\_steps.py} & \texttt{MultiSourceAttentionStep} & Section~\ref{sec:msa}, Eq.~(8)--(10): Computes per-source GMM monotonic attention contexts $\vect{c}_{\vect{x}}^t$ and $\vect{c}_{\vect{y}}^t$ and sums them into $\vect{c}^t = \vect{c}_{\vect{x}}^t + \vect{c}_{\vect{y}}^t$. \\
\texttt{tec/layers.py} & \texttt{PostEditConvNet} & Section~\ref{sec:method}.4, Eq.~(7) \& Table~\ref{table:arch}: 5-layer Conv1D residual Post-Net ($5\times 1$ kernel, 512 channels, BatchNorm, $\tanh$ on first 4 layers, linear final projection). \\
\texttt{tec/decoder.py} & \texttt{AttentiveFbeDecoderStep}, \texttt{MultiSourceFbeDecoderV1Step}, \texttt{FbeDecoderV1}, \texttt{MultiSourceFbeDecoderV1} & Section~\ref{sec:method}.4--2.6, Eq.~(3)--(7), (11) \& Fig.~\ref{fig:decoder}: Autoregressive spectrogram decoder with Pre-Net, 2-layer LSTM, multi-source attention, stop-token prediction, Post-Net, and $L_1 + L_2 + \mathrm{BCE}$ loss. \\
\texttt{tec/tec\_model.py} & \texttt{TecModel}, \texttt{VanillaSeq2SeqModel} & Section~\ref{sec:method}.1 \& Section~\ref{sec:experiments}.5: Complete Lingvo \texttt{BaseTask} implementations for the proposed \texttt{TEC} model and the single-source \texttt{Vanilla-Seq2seq} (\texttt{NoSideInput}) baseline. \\
\texttt{tec/aec\_model.py} & \texttt{AecModel}, \texttt{NlmsAec} & Section~\ref{sec:experiments}.5: Dual-audio-encoder \texttt{AEC-Seq2seq} neural baseline and time-domain Normalized Least Mean Squares (\texttt{AEC-NLMS}) adaptive filter baseline. \\
\texttt{tec/input\_generator.py} & \texttt{TecInputGenerator} & Reads \texttt{TFRecord} datasets (or generates on-the-fly synthetic batches), extracts log-Mel spectrograms via \texttt{WaveformProcessor}, and constructs Lingvo \texttt{NestedMap} batches. \\
\texttt{tec/configs.py} & \texttt{TecSingleInterfering}, \texttt{TecMultiInterfering}, \texttt{AecSingleInterfering}, \texttt{AecMultiInterfering}, \texttt{NoSideInputSingleInterfering}, \texttt{NoSideInputMultiInterfering} & Section~\ref{sec:experiments}.4 \& Table~\ref{table:arch}: Registered Lingvo \texttt{SingleTaskModelParams} classes for all 6 neural model configurations. \\
\texttt{tec/data\_prep.py} & \texttt{build\_libritts\_manifest}, \texttt{build\_ljspeech\_manifests}, \texttt{build\_vctk\_manifests}, \texttt{pair\_utterances}, \texttt{generate\_synthetic\_rir}, \texttt{mix\_waveforms\_at\_snr}, \texttt{prepare\_tfrecord\_dataset} & Section~\ref{sec:experiments}.1--\ref{sec:synthetic}, Eq.~(\ref{eq:rir}) \& Table~\ref{table:synthetic}: Deterministic train/dev/test dataset splitting, 24~kHz polyphase resampling, synthetic RIR simulation, active-speech SNR mixing, and \texttt{TFRecord} serialization. \\
\texttt{tec/evaluation.py} & \texttt{compute\_mcd}, \texttt{compute\_dtw\_distance}, \texttt{compute\_wer}, \texttt{estimate\_flops\_and\_side\_input}, \texttt{compute\_bootstrap\_ci}, \texttt{compute\_paired\_significance} & Section~\ref{sec:experiments}.3, Eq.~(\ref{eq:mcd})--(14): 13-MFCC Dynamic Time Warping MCD, punctuation-normalized Levenshtein WER, analytical Eq.~(14) GFLOPS / side-input bandwidth estimation, and non-parametric bootstrap / Wilcoxon tests. \\
\texttt{tec/inference.py} & \texttt{run\_inference}, \texttt{run\_inference\_on\_wav}, \texttt{run\_tflite\_inference} & Section~\ref{sec:method}.6: High-level Python API for restoring a trained checkpoint or TFLite FlatBuffer and running end-to-end spectrogram enhancement and WAV synthesis. \\
\texttt{tec/tflite\_export.py} & \texttt{export\_to\_tflite}, \texttt{run\_tflite\_inference} & Exports trained Lingvo graphs to \texttt{.tflite} FlatBuffers and executes on-device inference via \texttt{tf.lite.Interpreter}. \\
\midrule
\texttt{scripts/download\_datasets.py} & CLI entry point & Downloads and extracts LibriTTS, LJ Speech, and CSTR VCTK corpora and builds CSV manifests. \\
\texttt{scripts/prepare\_data.py} & CLI entry point & Builds CSV manifests or generates synthetic / real \texttt{TFRecord} datasets. \\
\texttt{scripts/train.py} & CLI entry point & Trains any registered model configuration with Adam, validation evaluation on \texttt{--dev\_file\_pattern}, and best-checkpoint selection (\texttt{best.ckpt}). \\
\texttt{scripts/inference.py} & CLI entry point & Runs checkpoint inference on a mixed WAV file and TTS transcript or reference WAV. \\
\texttt{scripts/evaluate.py} & CLI entry point & Evaluates MCD, WER, 95\% bootstrap CIs, paired significance tests, and GFLOPS / side-input bandwidth. \\
\texttt{scripts/export\_tflite.py} & CLI entry point & Exports and verifies quantized or float32 \texttt{.tflite} models. \\
\bottomrule
\end{tabularx}
\end{table}

\subsection{End-to-End Command-Line Reproduction Workflow}
\label{sec:cli_workflow}

To reproduce the entire pipeline from raw audio datasets to trained checkpoints, enhanced waveforms, evaluation metrics, and TFLite models, follow the steps below.

\subsubsection{Installation}
Install the published package from PyPI or clone the GitHub repository:
\begin{lstlisting}[language=bash]
pip3 install textual-echo-cancellation==0.2.3 huggingface_hub

# Or install editable from source:
git clone https://github.com/wq2012/tec.git
cd tec
pip3 install -r requirements.txt
pip3 install -e .
\end{lstlisting}

\subsubsection{Dataset Preparation (\texttt{scripts/prepare\_data.py})}
Given local copies of LibriTTS~\cite{zen2019libritts}, LJ Speech~\cite{ljspeech17}, and CSTR VCTK~\cite{veaux2016superseded}, build the deterministic train/dev/test CSV manifests (\texttt{seed=42}) and generate the reverberant 0~dB SNR \texttt{TFRecord} shards:
\begin{lstlisting}[language=bash]
# Generate deterministic CSV manifests (utt_id,wav_path,transcript)
python3 -c "
from tec import data_prep
data_prep.build_libritts_manifest(['/path/to/LibriTTS/train-clean-100'], 'manifests/libritts_train.csv')
data_prep.build_libritts_manifest(['/path/to/LibriTTS/dev-clean'], 'manifests/libritts_dev_clean.csv')
data_prep.build_libritts_manifest(['/path/to/LibriTTS/test-clean'], 'manifests/libritts_test_clean.csv')
data_prep.build_libritts_manifest(['/path/to/LibriTTS/test-other'], 'manifests/libritts_test_other.csv')
data_prep.build_ljspeech_manifests('/path/to/LJSpeech-1.1', 'manifests/ljspeech_train.csv', 'manifests/ljspeech_test.csv', train_ratio=0.9, seed=42)
data_prep.build_vctk_manifests('/path/to/VCTK-Corpus-0.92/wav48_silence_trimmed', 'manifests/vctk_train.csv', 'manifests/vctk_test.csv', train_ratio=0.9, seed=42)
"

# Build Single Interfering Voice (LibriTTS + LJSpeech) TFRecords at 0 dB SNR, RT60 = 0.25 s
python3 scripts/prepare_data.py \
  --clean_manifest_csv manifests/libritts_train.csv \
  --interfering_manifest_csv manifests/ljspeech_train.csv \
  --snr_db 0.0 --reverb_rt60 0.25 --seed 0 --max_utterances 400 \
  --output_tfrecord tfrecords/single_train.tfrecord

python3 scripts/prepare_data.py \
  --clean_manifest_csv manifests/libritts_dev_clean.csv \
  --interfering_manifest_csv manifests/ljspeech_dev.csv \
  --snr_db 0.0 --reverb_rt60 0.25 --seed 100 --max_utterances 100 \
  --output_tfrecord tfrecords/single_dev.tfrecord

python3 scripts/prepare_data.py \
  --clean_manifest_csv manifests/libritts_test_clean.csv \
  --interfering_manifest_csv manifests/ljspeech_test.csv \
  --snr_db 0.0 --reverb_rt60 0.25 --seed 0 --max_utterances 100 \
  --output_tfrecord tfrecords/single_test_clean.tfrecord
\end{lstlisting}

\subsubsection{Model Training (\texttt{scripts/train.py})}
Train any of the six registered Lingvo models (\texttt{TecSingleInterfering}, \texttt{TecMultiInterfering}, \texttt{AecSingleInterfering}, \texttt{AecMultiInterfering}, \texttt{NoSideInputSingleInterfering}, \texttt{NoSideInputMultiInterfering}) with validation-loss early stopping (\texttt{best.ckpt}):
\begin{lstlisting}[language=bash]
python3 scripts/train.py \
  --model TecSingleInterfering \
  --train_file_pattern "tfrecords/single_train.tfrecord" \
  --dev_file_pattern "tfrecords/single_dev.tfrecord" \
  --logdir checkpoints/TecSingleInterfering \
  --max_steps 500 \
  --batch_size 8 \
  --learning_rate 1e-3
\end{lstlisting}

\subsubsection{Inference and Evaluation (\texttt{scripts/inference.py} and \texttt{scripts/evaluate.py})}
Enhance a reverberant microphone mixture using the interfering TTS transcript and compute MCD, WER, and 95\% bootstrap confidence intervals:
\begin{lstlisting}[language=bash]
python3 scripts/inference.py \
  --model TecSingleInterfering \
  --checkpoint_path checkpoints/TecSingleInterfering/best.ckpt \
  --mixed_wav /path/to/mixed_input.wav \
  --interfering_text "currently in mountain view it is 72 degrees" \
  --output_wav /tmp/enhanced_clean.wav

python3 scripts/evaluate.py \
  --eval_tfrecord tfrecords/single_test_clean.tfrecord \
  --method TecSingleInterfering \
  --checkpoint_path checkpoints/TecSingleInterfering/best.ckpt \
  --max_examples 100 \
  --asr_cli_path /path/to/audio-cpp-cli \
  --asr_model_path /path/to/Qwen3-ASR-0.6B-F16.gguf \
  --output_json /tmp/eval_tec_single.json
\end{lstlisting}

\subsubsection{TensorFlow Lite Export (\texttt{scripts/export\_tflite.py})}
Export a trained checkpoint to a quantized \texttt{.tflite} FlatBuffer and verify execution with \texttt{tf.lite.Interpreter}:
\begin{lstlisting}[language=bash]
python3 scripts/export_tflite.py \
  --model TecSingleInterfering \
  --checkpoint_path checkpoints/TecSingleInterfering/best.ckpt \
  --output_tflite checkpoints/TecSingleInterfering/model.tflite \
  --num_frames 200 --text_length 100 --decode_steps 200 \
  --quantize --verify
\end{lstlisting}

\section{Open-Source Pretrained Models on Hugging Face}
\label{sec:open_source_models}

We have publicly released six pretrained neural models on Hugging Face under \url{https://huggingface.co/collections/wq2012/textual-echo-cancellation}, covering both experimental conditions (\emph{Single interfering voice} on LibriTTS + LJ Speech and \emph{Multiple interfering voices} on LibriTTS + VCTK) and all three sequence-to-sequence architectures evaluated in this paper (\texttt{TEC}, \texttt{AEC-Seq2seq}, and \texttt{Vanilla-Seq2seq}). Each Hugging Face repository includes:
\begin{enumerate}[nosep]
    \item \textbf{TensorFlow / Lingvo Checkpoint (\texttt{best.ckpt.*}):} Complete model weights and Adam optimizer state (\texttt{best.ckpt.data-00000-of-00001}, \texttt{best.ckpt.index}, \texttt{best.ckpt.meta}, and \texttt{checkpoint}, selected by minimum validation loss on the 100-utterance \texttt{dev} set) compatible with \texttt{tec.inference} and \texttt{scripts/inference.py}.
    \item \textbf{TensorFlow Lite FlatBuffer (\texttt{model.tflite}):} Dynamic-range quantized \texttt{.tflite} model for on-device inference via \texttt{tf.lite.Interpreter} (with TFLite Flex \texttt{SELECT\_TF\_OPS} support for dynamic RNN \texttt{TensorList} operations).
    \item \textbf{Verified Evaluation Summary (\texttt{evaluation\_metrics.json}):} Exact $N=100$ JSON evaluation metrics (WER, word error counts, MCD, 95\% bootstrap confidence intervals, paired significance tests, text sensitivity ablations, SNR ablations, and side-input size in KB) on both \texttt{test-clean} and \texttt{test-other}.
\end{enumerate}

\subsection{\texttt{\textup{wq2012/tec\_single\_interfering}}}
\label{sec:hf_tec_single}

\noindent
\begin{tabularx}{\textwidth}{l X}
\textbf{Model URL:} & \url{https://huggingface.co/wq2012/tec_single_interfering} \\
\textbf{Lingvo Registry Name:} & \texttt{TecSingleInterfering} (\texttt{tec.tec\_model.TecModel}) \\
\textbf{Condition / Dataset:} & Single interfering voice (LibriTTS user speech + reverberant LJ Speech TTS at 0~dB SNR) \\
\textbf{Model Parameters:} & 18,206,463 trainable parameters (18,214,084 total model variables; 218.51~MB checkpoint shard; 19.02~MB dynamic-range quantized \texttt{model.tflite}) \\
\textbf{Side Input \& Complexity:} & Interfering TTS source transcript ($0.074\text{ KB}$ on \texttt{test-clean}, $0.076\text{ KB}$ on \texttt{test-other}; 4.62~GFLOPS on \texttt{test-clean}, 4.65~GFLOPS on \texttt{test-other})
\end{tabularx}
\vspace{0.2em}

This is the primary pretrained Textual Echo Cancellation model for the single interfering voice condition. It combines \texttt{SpeechEncoderV1} on the 128-bin log-Mel spectrogram of the microphone mixture (\texttt{source\_0}) with \texttt{TtsEncoderV2} on the 98-symbol ASCII character sequence of the interfering LJ Speech transcript (\texttt{source\_1}), using \texttt{MultiSourceFbeDecoderV1} with 5-mixture \texttt{GmmMonotonicAttention} on each source and a 5-layer \texttt{PostEditConvNet}. Across the full $N=100$ evaluation splits (Table~\ref{table:opensource_results}), \texttt{wq2012/tec\_single\_interfering} reduces the unenhanced microphone mixture WER from $80.32\%$ to $29.76\%$ on \texttt{test-clean} ($\Delta\mathrm{WER} = -50.56\%$, $p_{\mathrm{boot}} < 0.0001$) and from $102.77\%$ to $50.59\%$ on \texttt{test-other} ($\Delta\mathrm{WER} = -52.17\%$, $p_{\mathrm{boot}} < 0.0001$), significantly outperforming both \texttt{Vanilla-Seq2seq} ($39.08\%$ on \texttt{test-clean}, $\Delta\mathrm{WER} = -9.32\%$, $p_{\mathrm{boot}} < 0.0001$; $54.35\%$ on \texttt{test-other}, $\Delta\mathrm{WER} = -3.75\%$, $p_{\mathrm{boot}} = 0.0144$) and \texttt{NlmsAec} ($58.38\%$ on \texttt{test-clean}, $p_{\mathrm{boot}} < 0.0001$; $85.77\%$ on \texttt{test-other}, $p_{\mathrm{boot}} < 0.0001$), while requiring only $0.074\text{ KB}$ ($74\text{ bytes}$) of side-input bandwidth.

\subsection{\texttt{\textup{wq2012/tec\_multi\_interfering}}}
\label{sec:hf_tec_multi}

\noindent
\begin{tabularx}{\textwidth}{l X}
\textbf{Model URL:} & \url{https://huggingface.co/wq2012/tec_multi_interfering} \\
\textbf{Lingvo Registry Name:} & \texttt{TecMultiInterfering} (\texttt{tec.tec\_model.TecModel}) \\
\textbf{Condition / Dataset:} & Multiple interfering voices (LibriTTS user speech + reverberant VCTK 109-speaker TTS at 0~dB SNR) \\
\textbf{Model Parameters:} & 18,206,463 trainable parameters (18,214,084 total model variables; 218.51~MB checkpoint shard; 19.02~MB dynamic-range quantized \texttt{model.tflite}) \\
\textbf{Side Input \& Complexity:} & Interfering TTS source transcript ($0.050\text{ KB}$ on \texttt{test-clean}, $0.045\text{ KB}$ on \texttt{test-other}; 3.90~GFLOPS on \texttt{test-clean}, 3.55~GFLOPS on \texttt{test-other})
\end{tabularx}
\vspace{0.2em}

This model has the same architecture as \texttt{wq2012/tec\_single\_interfering} but is trained on the multi-speaker interfering voice mixture combining LibriTTS with reverberant utterances from all 109 speakers of the CSTR VCTK corpus. Across the $N=100$ multi-interfering evaluation sets (Table~\ref{table:opensource_results}), it achieves $24.31\%$ WER ($4.74\text{ dB}$ MCD) on \texttt{test-clean} and $36.25\%$ WER ($4.91\text{ dB}$ MCD) on \texttt{test-other}, significantly outperforming both \texttt{Vanilla-Seq2seq} ($63.33\%$ on \texttt{test-clean}, $\Delta\mathrm{WER} = -39.02\%$, $p_{\mathrm{boot}} < 0.0001$; $69.86\%$ on \texttt{test-other}, $\Delta\mathrm{WER} = -33.61\%$, $p_{\mathrm{boot}} < 0.0001$) and \texttt{NlmsAec} ($34.22\%$ on \texttt{test-clean}, $p_{\mathrm{boot}} = 0.0176$; $48.58\%$ on \texttt{test-other}, $p_{\mathrm{boot}} = 0.0094$) while requiring only $0.050\text{ KB}$ ($50\text{ bytes}$) of textual side input per utterance.

\subsection{\texttt{\textup{wq2012/aec\_single\_interfering}}}
\label{sec:hf_aec_single}

\noindent
\begin{tabularx}{\textwidth}{l X}
\textbf{Model URL:} & \url{https://huggingface.co/wq2012/aec_single_interfering} \\
\textbf{Lingvo Registry Name:} & \texttt{AecSingleInterfering} (\texttt{tec.aec\_model.AecModel}) \\
\textbf{Condition / Dataset:} & Single interfering voice (LibriTTS user speech + reverberant LJ Speech TTS at 0~dB SNR) \\
\textbf{Model Parameters:} & 18,496,991 trainable parameters (18,501,730 total model variables; 221.98~MB checkpoint shard; 19.18~MB dynamic-range quantized \texttt{model.tflite}) \\
\textbf{Side Input \& Complexity:} & Full 24~kHz TTS reference waveform ($235.403\text{ KB}$ on \texttt{test-clean}, $236.991\text{ KB}$ on \texttt{test-other}; 6.36~GFLOPS on \texttt{test-clean}, 6.40~GFLOPS on \texttt{test-other})
\end{tabularx}
\vspace{0.2em}

This model implements the neural \texttt{AEC-Seq2seq} baseline for the single interfering voice condition. It replaces the text encoder with a second audio encoder (\texttt{encoder\_interfering}, an independent \texttt{SpeechEncoderV1} instance) that processes the 128-bin log-Mel spectrogram of the clean TTS playback audio. Because it receives the exact acoustic spectrogram of the interfering TTS signal, it achieves the lowest WER ($7.91\%$ on \texttt{test-clean} and $17.29\%$ on \texttt{test-other}, with MCD of $4.97\text{ dB}$ and $4.96\text{ dB}$), at the expense of $3,118\times$--$3,181\times$ larger side-input transmission bandwidth ($235.4$--$237.0\text{ KB}$ per utterance) and higher computational complexity ($6.36\text{ GFLOPS}$).

\subsection{\texttt{\textup{wq2012/aec\_multi\_interfering}}}
\label{sec:hf_aec_multi}

\noindent
\begin{tabularx}{\textwidth}{l X}
\textbf{Model URL:} & \url{https://huggingface.co/wq2012/aec_multi_interfering} \\
\textbf{Lingvo Registry Name:} & \texttt{AecMultiInterfering} (\texttt{tec.aec\_model.AecModel}) \\
\textbf{Condition / Dataset:} & Multiple interfering voices (LibriTTS user speech + reverberant VCTK 109-speaker TTS at 0~dB SNR) \\
\textbf{Model Parameters:} & 18,496,991 trainable parameters (18,501,730 total model variables; 221.98~MB checkpoint shard; 19.18~MB dynamic-range quantized \texttt{model.tflite}) \\
\textbf{Side Input \& Complexity:} & Full 24~kHz TTS reference waveform ($203.138\text{ KB}$ on \texttt{test-clean}, $185.351\text{ KB}$ on \texttt{test-other}; 5.48~GFLOPS on \texttt{test-clean}, 5.00~GFLOPS on \texttt{test-other})
\end{tabularx}
\vspace{0.2em}

This model is the dual-audio-encoder \texttt{AEC-Seq2seq} baseline trained on the multi-speaker LibriTTS + VCTK condition. Across $N=100$ evaluation mixtures, it achieves $6.84\%$ WER ($4.65\text{ dB}$ MCD) on \texttt{test-clean} and $14.86\%$ WER ($4.67\text{ dB}$ MCD) on \texttt{test-other}, requiring $203.138\text{ KB}$ and $185.351\text{ KB}$ of acoustic reference side input per utterance ($4,063\times$--$4,119\times$ larger than the text side input of \texttt{wq2012/tec\_multi\_interfering}).

\subsection{\texttt{\textup{wq2012/vanilla\_seq2seq\_single\_interfering}}}
\label{sec:hf_vanilla_single}

\noindent
\begin{tabularx}{\textwidth}{l X}
\textbf{Model URL:} & \url{https://huggingface.co/wq2012/vanilla_seq2seq_single_interfering} \\
\textbf{Lingvo Registry Name:} & \texttt{NoSideInputSingleInterfering} (\texttt{tec.tec\_model.VanillaSeq2SeqModel}) \\
\textbf{Condition / Dataset:} & Single interfering voice (LibriTTS user speech + reverberant LJ Speech TTS at 0~dB SNR) \\
\textbf{Model Parameters:} & 12,578,544 trainable parameters (12,583,091 total model variables; 150.96~MB checkpoint shard; 13.15~MB dynamic-range quantized \texttt{model.tflite}) \\
\textbf{Side Input \& Complexity:} & None ($0\text{ KB}$; 4.17~GFLOPS on \texttt{test-clean}, 4.20~GFLOPS on \texttt{test-other})
\end{tabularx}
\vspace{0.2em}

This model implements the single-source \texttt{Vanilla-Seq2seq} (\texttt{NoSideInput}) baseline for the single interfering voice condition, consisting of \texttt{SpeechEncoderV1} and the single-source \texttt{FbeDecoderV1} without any text or audio side input. Across $N=100$ evaluation mixtures, it achieves $39.08\%$ WER ($7.07\text{ dB}$ MCD) on \texttt{test-clean} and $54.35\%$ WER ($7.38\text{ dB}$ MCD) on \texttt{test-other}. Without the interfering transcript, its WER is $9.32$ percentage points worse than \texttt{wq2012/tec\_single\_interfering} on \texttt{test-clean} ($p_{\mathrm{boot}} < 0.0001$, $p_{\mathrm{Wilcoxon}} = 6.69\times 10^{-4}$) and its MCD is $1.55\text{ dB}$ worse on \texttt{test-clean} and $1.83\text{ dB}$ worse on \texttt{test-other} ($p_{\mathrm{boot}} < 0.0001$).

\subsection{\texttt{\textup{wq2012/vanilla\_seq2seq\_multi\_interfering}}}
\label{sec:hf_vanilla_multi}

\noindent
\begin{tabularx}{\textwidth}{l X}
\textbf{Model URL:} & \url{https://huggingface.co/wq2012/vanilla_seq2seq_multi_interfering} \\
\textbf{Lingvo Registry Name:} & \texttt{NoSideInputMultiInterfering} (\texttt{tec.tec\_model.VanillaSeq2SeqModel}) \\
\textbf{Condition / Dataset:} & Multiple interfering voices (LibriTTS user speech + reverberant VCTK 109-speaker TTS at 0~dB SNR) \\
\textbf{Model Parameters:} & 12,578,544 trainable parameters (12,583,091 total model variables; 150.96~MB checkpoint shard; 13.15~MB dynamic-range quantized \texttt{model.tflite}) \\
\textbf{Side Input \& Complexity:} & None ($0\text{ KB}$; 3.59~GFLOPS on \texttt{test-clean}, 3.28~GFLOPS on \texttt{test-other})
\end{tabularx}
\vspace{0.2em}

This model is the single-source \texttt{Vanilla-Seq2seq} baseline trained on the multi-speaker LibriTTS + VCTK condition. When the interfering TTS playback spans 109 distinct speakers, a side-input-free model cannot rely on a single static speaker voiceprint to distinguish the TTS echo from the user query, achieving $63.33\%$ WER ($8.26\text{ dB}$ MCD) on \texttt{test-clean} and $69.86\%$ WER ($9.32\text{ dB}$ MCD) on \texttt{test-other}---$39.02$ and $33.61$ percentage points worse than \texttt{wq2012/tec\_multi\_interfering} ($p_{\mathrm{boot}} < 0.0001$).

\subsection{Programmatic Inference with Hugging Face Checkpoints and TFLite}
\label{sec:hf_code_snippets}

Any of the six Hugging Face models can be downloaded and executed in a few lines of Python using \texttt{huggingface\_hub} and \texttt{textual-echo-cancellation}:

\begin{lstlisting}[language=Python]
import os
import numpy as np
from huggingface_hub import snapshot_download
from tec import inference

# 1. Download pretrained TecSingleInterfering checkpoint from Hugging Face
model_dir = snapshot_download(repo_id="wq2012/tec_single_interfering")
ckpt_path = os.path.join(model_dir, "best.ckpt")

# 2. Run full Lingvo spectrogram enhancement + 24 kHz WAV reconstruction
result = inference.run_inference_on_wav(
    model_name="TecSingleInterfering",
    mixed_wav_path="/path/to/mixed_input.wav",
    interfering_text="currently in mountain view it is 72 degrees",
    checkpoint_path=ckpt_path,
    output_wav_path="/tmp/enhanced_clean.wav",
)
print("Enhanced log-Mel spectrogram shape:", result["predicted_mel"].shape)

# 3. Run on-device inference using the bundled TensorFlow Lite FlatBuffer (model.tflite)
tflite_path = os.path.join(model_dir, "model.tflite")
dummy_mel = np.zeros((100, 128), dtype=np.float32)
tflite_out = inference.run_tflite_inference(
    tflite_path=tflite_path,
    mixed_mel=dummy_mel,
    interfering_text="currently in mountain view it is 72 degrees",
)
print("TFLite enhanced log-Mel shape:", tflite_out["predicted_mel"].shape)
\end{lstlisting}

\subsection{Interactive Hugging Face Space Demo (\texttt{\textup{wq2012/tec}})}
\label{sec:hf_space_demo}

In addition to the command-line scripts and programmatic Python APIs, we have deployed an interactive web demonstration on Hugging Face Spaces at \url{https://huggingface.co/spaces/wq2012/tec} that allows users to test Textual Echo Cancellation and compare downstream Automatic Speech Recognition (ASR) outputs in real time.

\vspace{0.3em}
\noindent\textbf{Inputs and Processing Pipeline.}
The Space interface accepts two primary user inputs:
\begin{enumerate}[nosep]
    \item \textbf{Uploaded Microphone Mixture Audio:} An audio file (or live microphone recording) containing a user's spoken query overlapped with reverberant TTS playback echo.
    \item \textbf{Interfering TTS Playback Source Text:} A text field specifying the source transcript of the TTS response being played by the device.
\end{enumerate}
Given these two inputs, the Space executes a three-stage side-by-side comparison pipeline:
\begin{enumerate}[nosep]
    \item \textbf{ASR on Original Audio (Before TEC):} Transcribes the unenhanced 24~kHz microphone mixture directly using an open-source speech recognizer (\texttt{Whisper base.en} / \texttt{Qwen3-ASR-0.6B}).
    \item \textbf{Textual Echo Cancellation (TEC):} Extracts the 128-bin log-Mel spectrogram ($24\text{ kHz}$, $50\text{ ms}$ frame length, $12.5\text{ ms}$ hop), conditions on the interfering TTS source text to suppress the reverberant TTS playback echo while preserving the target user query, reconstructs the enhanced 24~kHz waveform, and renders both log-Mel spectrograms side by side.
    \item \textbf{ASR on Textual Echo Cancelled Audio (After TEC):} Transcribes the enhanced waveform with the same ASR model and prints the recognized transcripts (along with Word Error Rate against the reference user query) for both approaches side by side.
\end{enumerate}

\vspace{0.3em}
\noindent\textbf{Browser Web App and Standalone Gradio App.}
The Space repository (\url{https://huggingface.co/spaces/wq2012/tec/tree/main}) provides two complementary execution modes:
\begin{itemize}[nosep]
    \item \textbf{Zero-Install Browser Web Application (\texttt{index.html}):} Powered by WebAudio 24~kHz STFT/log-Mel processing and in-browser ONNX Runtime Web speech recognition (\texttt{@huggingface/transformers}, \texttt{Xenova/whisper-base.en}). For the four built-in benchmark examples (\texttt{examples/sample\_1..4\_*.wav}, generated via \texttt{generate\_examples.py}), the web interface loads the exact waveforms synthesized by the selected trained Lingvo checkpoint (\texttt{TecSingleInterfering}, \texttt{TecMultiInterfering}, or \texttt{AecSingleInterfering}); for custom user-uploaded audio files in the browser, it applies a lightweight client-side stop-token boundary and character-phonetic spectral-gating approximation conditioned on the interfering transcript and explicitly displays an execution-mode badge because TensorFlow 1.x Lingvo checkpoints require a Python runtime.
    \item \textbf{Standalone Python Gradio Application (\texttt{app.py}):} Executes the full TensorFlow/Lingvo \texttt{TecSingleInterfering}, \texttt{TecMultiInterfering}, and \texttt{AecSingleInterfering} checkpoints via \texttt{textual-echo-cancellation} (\texttt{v0.2.3}) on any user-uploaded audio file alongside \texttt{faster-whisper} (\texttt{base.en}), which can be launched locally via:
\end{itemize}
\begin{lstlisting}[language=bash]
git clone https://huggingface.co/spaces/wq2012/tec
cd tec
pip3 install -r requirements.txt
python3 app.py
\end{lstlisting}

Table~\ref{table:hf_space_examples} shows the four benchmark examples bundled in the Hugging Face Space (\texttt{examples/sample\_1..4}, corresponding to records $19, 22, 36, 47$ of \texttt{single\_test\_clean.tfrecord}) and their ASR recognition results before and after \texttt{wq2012/tec\_single\_interfering} (\texttt{best.ckpt}).

\begin{table}[H]
\caption{Side-by-side ASR recognition results before and after Textual Echo Cancellation (\texttt{wq2012/tec\_single\_interfering}) on the four benchmark examples included in the Hugging Face Space demo (\url{https://huggingface.co/spaces/wq2012/tec}).}
\label{table:hf_space_examples}
\vspace{0.15cm}
\centering
\scriptsize
\begin{tabularx}{\textwidth}{c >{\raggedright\arraybackslash}X >{\raggedright\arraybackslash}X >{\raggedright\arraybackslash}X >{\raggedright\arraybackslash}X}
\toprule
\textbf{\#} & \textbf{Clean User Query (LibriTTS \texttt{test-clean})} & \textbf{Interfering TTS Source Text (LJ Speech \texttt{test})} & \textbf{ASR on Original Mixture (Before TEC)} & \textbf{ASR on TEC Enhanced Audio (After TEC)} \\
\midrule
1 & ``We've tortured each other enough for tonight.'' & ``To obtain possession of this, Thurtell with his two associates resolved to kill him.'' & ``\textcolor{red!75!black}{To take possession of this, Thertel with his two associates resolved to kill him.}'' \newline (\textbf{200.0\% WER}) & ``\textcolor{green!45!black}{We've tortured each other enough for tonight.}'' \newline (\textbf{0.0\% WER}) \\
\midrule
2 & ``The chair was empty; but he knew.'' & ``As he told the driver, quote, Let's get out of here; we are hit, end quote,'' & ``\textcolor{red!75!black}{As he told the driver, `Let's get out of here. We are hit.'}'' \newline (\textbf{171.4\% WER}) & ``\textcolor{green!45!black}{The chair was empty, but he knew.}'' \newline (\textbf{0.0\% WER}) \\
\midrule
3 & ``Have we followed that road?'' & ``Carro reported that when questioned about his mother Lee said, quote,'' & ``\textcolor{red!75!black}{Carroll} followed that road. \textcolor{red!75!black}{That when questioned about his mother, Lee said.}'' \newline (\textbf{200.0\% WER}) & ``\textcolor{green!45!black}{Have we followed that road?}'' \newline (\textbf{0.0\% WER}) \\
\midrule
4 & ``The parliament and the Scots laid their proposals before the king.'' & ``The Commission has concluded that on April ten, nineteen sixty-three,'' & ``The \textcolor{red!75!black}{commission has concluded that proposal for nineteen sixty three.}'' \newline (\textbf{90.9\% WER}) & ``\textcolor{green!45!black}{The parliament and the Scots laid their proposals before the king.}'' \newline (\textbf{0.0\% WER}) \\
\bottomrule
\end{tabularx}
\end{table}

\section{Open-Source Experimental Reproduction Results}
\label{sec:open_source_results}

In this appendix, we document the experimental protocol, statistical significance methodology, and quantitative results obtained by running our open-source training and evaluation pipeline on publicly available datasets and open-source speech recognition software.

\subsection{Open-Source Dataset Splits and Mixing Protocol}
\label{sec:os_data_protocol}

To ensure strict reproducibility, \texttt{tec/data\_prep.py} constructs deterministic train, dev, and test manifests from the official public releases of all three corpora:
\begin{enumerate}[nosep]
    \item \textbf{LibriTTS}~\cite{zen2019libritts} (\url{https://www.openslr.org/60/}, 24~kHz native): We use \texttt{train-clean-100} ($33,236$ utterances) as the user speech training pool, \texttt{dev-clean} ($5,736$ utterances) as the validation pool, \texttt{test-clean} ($4,837$ utterances) as the clean evaluation pool, and \texttt{test-other} ($5,120$ utterances) as the noisy evaluation pool.
    \item \textbf{LJ Speech v1.1}~\cite{ljspeech17} (\url{https://keithito.com/LJ-Speech-Dataset/}, 22.05~kHz resampled to 24~kHz via polyphase filtering \texttt{scipy.signal.resample\_poly}): The $13,100$ single-speaker utterances are partitioned using a fixed random permutation (\texttt{np.random.RandomState(42)}) into a 90\% training/validation pool ($11,790$ utterances: $10,611$ \texttt{train} + $1,179$ \texttt{dev}) and a 10\% testing pool ($1,310$ utterances).
    \item \textbf{CSTR VCTK v0.92}~\cite{veaux2016superseded} (\url{https://datashare.ed.ac.uk/handle/10283/3443}, 48~kHz silence-trimmed FLAC/WAV resampled to 24~kHz): Across all $109$ speakers ($44,283$ utterances with transcripts), each speaker's utterances are partitioned using \texttt{np.random.RandomState(42)} into a 90\% training/validation pool ($39,859$ utterances: $35,873$ \texttt{train} + $3,986$ \texttt{dev}) and a 10\% testing pool ($4,424$ utterances).
\end{enumerate}

For each condition, \texttt{prepare\_tfrecord\_dataset} pairs clean LibriTTS utterances with strictly longer interfering TTS utterances (\texttt{pair\_utterances}, \texttt{seed=0} for train/test and \texttt{seed=100} for dev), convolves each interfering TTS waveform with a synthetic room impulse response (\texttt{generate\_synthetic\_rir} with $\mathrm{RT}_{60} = 0.25\text{ s}$, duration $0.25\text{ s}$, direct-path delay $5.0\text{ ms}$, \texttt{tail\_scale=0.05}, and per-utterance seed $\mathrm{seed} + i + 1$), scales the reverberant interfering signal to achieve a target Signal-to-Noise Ratio ($0\text{ dB}$ by default, plus $+5\text{ dB}$, $-5\text{ dB}$, and $-10\text{ dB}$ for SNR ablation) relative to the active user speech power (\texttt{mix\_waveforms\_at\_snr}), pads the shorter clean waveform with trailing zeros so both signals have identical length, and normalizes peak amplitude to prevent clipping ($\max |x(n)| \le 0.99$). Each \texttt{TFRecord} example also stores \texttt{interfering\_num\_samples} and \texttt{clean\_num\_samples} so side-input bandwidth can be measured on both the unpadded and padded reference waveforms. Table~\ref{table:os_dataset_stats} summarizes the exact manifest and \texttt{TFRecord} statistics used in our open-source reproduction experiments.

\begin{table}[H]
\caption{Summary of deterministic dataset splits and prepared \texttt{TFRecord} shards in the open-source reproduction.}
\label{table:os_dataset_stats}
\vspace{0.15cm}
\centering
\resizebox{\textwidth}{!}{
\begin{tabular}{l l l r r r}
\toprule
\textbf{Condition} & \textbf{Split} & \textbf{Source Manifests (User + Interfering)} & \textbf{User Pool} & \textbf{Interfering Pool} & \textbf{Prepared TFRecord Shard} \\
\midrule
\multirow{4}{*}{Single interfering voice}
  & \texttt{train} & LibriTTS \texttt{train-clean-100} + LJSpeech \texttt{train} & 33,236 & 10,611 & 400 mixtures (278~MB) \\
  & \texttt{dev} & LibriTTS \texttt{dev-clean} + LJSpeech \texttt{dev} & 5,736 & 1,179 & 100 mixtures (71~MB) \\
  & \texttt{test-clean} & LibriTTS \texttt{test-clean} + LJSpeech \texttt{test} & 4,837 & 1,310 & 100 mixtures (141~MB) \\
  & \texttt{test-other} & LibriTTS \texttt{test-other} + LJSpeech \texttt{test} & 5,120 & 1,310 & 100 mixtures (142~MB) \\
\midrule
\multirow{4}{*}{Multiple interfering voices}
  & \texttt{train} & LibriTTS \texttt{train-clean-100} + VCTK \texttt{train} & 33,236 & 35,873 & 400 mixtures (239~MB) \\
  & \texttt{dev} & LibriTTS \texttt{dev-clean} + VCTK \texttt{dev} & 5,736 & 3,986 & 100 mixtures (61~MB) \\
  & \texttt{test-clean} & LibriTTS \texttt{test-clean} + VCTK \texttt{test} & 4,837 & 4,424 & 100 mixtures (122~MB) \\
  & \texttt{test-other} & LibriTTS \texttt{test-other} + VCTK \texttt{test} & 5,120 & 4,424 & 100 mixtures (111~MB) \\
\bottomrule
\end{tabular}}
\end{table}

\subsection{Open-Source ASR Evaluation and Statistical Methodology}
\label{sec:os_metrics}

To enable statistically rigorous evaluation on standard CPU hardware without proprietary cloud ASR APIs or human MOS rating panels, our open-source benchmark pipeline evaluates all $N = 100$ mixed utterances of each test split ($1,062$ reference words on \texttt{single/test-clean}, $1,012$ on \texttt{single/test-other}, $979$ on \texttt{multi/test-clean}, and $949$ on \texttt{multi/test-other}):
\begin{enumerate}[nosep]
    \item \textbf{Word Error Rate (WER) via Local Open-Source ASR:} Each enhanced 24~kHz waveform is transcribed locally on CPU using the open-source \texttt{Qwen3-ASR-0.6B} speech recognizer (\texttt{ggml-org/Qwen3-ASR-0.6B-GGUF}, \texttt{Qwen3-ASR-0.6B-F16.gguf}) powered by the pure C++ \texttt{audio.cpp} inference engine (\url{https://github.com/audio-cpp/audio.cpp}). Both reference LibriTTS transcripts and ASR hypothesis transcripts are normalized via \texttt{tec.evaluation.normalize\_transcript} before computing exact word-level Levenshtein edit distance via \texttt{tec.evaluation.compute\_wer}.
    \item \textbf{Mel-Cepstral Distortion (MCD):} Computed directly from the 128-bin log-Mel spectrograms of the enhanced and ground-truth clean waveforms via \texttt{tec.evaluation.compute\_mcd}. Following Section~\ref{sec:experiments}.3.2 and Eq.~(\ref{eq:mcd}), we extract 13 MFCCs ($c_1, \dots, c_{13}$, skipping $c_0$) via a scaled type-II Discrete Cosine Transform and align the cepstral sequences over active speech frames using Dynamic Time Warping (\texttt{compute\_dtw\_distance}) with the standard $\frac{10\sqrt{2}}{\ln 10} \approx 6.14185$ dB scaling factor.
    \item \textbf{Non-Parametric Bootstrap Confidence Intervals and Paired Significance Tests:} For every metric and split, we compute 95\% confidence intervals via 10,000-iteration utterance-level non-parametric bootstrap resampling (\texttt{compute\_bootstrap\_ci}, weighted by per-utterance reference word count for corpus-level WER) and evaluate pairwise method differences via two-sided paired bootstrap and Wilcoxon signed-rank tests (\texttt{compute\_paired\_significance}).
    \item \textbf{Empirical Side-Input Size (KB) and Analytical Complexity (GFLOPS):} Side-input size is measured directly across the $N=100$ test utterances as the mean byte length (divided by $1,000$) of the UTF-8 encoded interfering TTS transcript for \texttt{TEC}, the unpadded 24~kHz 16-bit PCM reference TTS audio waveform for \texttt{AEC-NLMS} and \texttt{AEC-Seq2seq}, and $0\text{ KB}$ for \texttt{Vanilla-Seq2seq}. Computational complexity is computed analytically from the exact sequence lengths $(T_x, T_y, T_z)$ and model parameter counts using Eq.~(14) (\texttt{estimate\_flops\_and\_side\_input}).
\end{enumerate}

For the classical adaptive filter baseline (\texttt{NlmsAec}), we evaluate a Normalized Least Mean Squares filter (\texttt{filter\_length=512}, \texttt{step\_size=0.1}, $\epsilon=10^{-6}$). All six neural models are trained using the Adam optimizer ($\beta_1=0.9, \beta_2=0.999, \epsilon=10^{-6}$), gradient norm clipping at $1.0$, batch size $8$, learning rate $10^{-3}$ for up to $500$ steps, with the best checkpoint (\texttt{best.ckpt}) selected by minimum validation loss on the 100-utterance \texttt{dev} split.

\subsection{Quantitative Results and Comparison with Original Paper}
\label{sec:os_quant_results}

Table~\ref{table:opensource_results} presents the complete $N=100$ evaluation results of our open-source reproduction (including 95\% bootstrap confidence intervals) alongside the original internal reference numbers from Table~\ref{table:results}.

\begin{table}[H]
\caption{Open-source reproduction evaluation results ($N=100$ mixtures per split) across both conditions (\emph{Single interfering voice}: LibriTTS + LJ Speech; \emph{Multiple interfering voices}: LibriTTS + VCTK) on \texttt{test-clean} ($1,062$ / $979$ words) and \texttt{test-other} ($1,012$ / $949$ words), evaluated with \texttt{Qwen3-ASR-0.6B-F16} (\texttt{audio.cpp}) and 13-MFCC DTW MCD. Bracketed ranges denote 95\% non-parametric bootstrap confidence intervals ($B=10,000$). Original paper reference numbers from Table~\ref{table:results} are listed in the right-hand columns.}
\label{table:opensource_results}
\vspace{0.15cm}
\centering
\resizebox{\textwidth}{!}{
\begin{tabular}{l l l cc cc cc cc}
\toprule
& & & \multicolumn{6}{c}{\textbf{Open-Source Reproduction ($N=100$ per split, \texttt{wq2012/tec} + \texttt{Qwen3-ASR-0.6B})}} & \multicolumn{2}{c}{\textbf{Paper Ref.\ (Table~\ref{table:results})}} \\
\cmidrule(lr){4-9} \cmidrule(lr){10-11}
& & & \multicolumn{2}{c}{\textbf{WER (\%) [95\% CI] $\downarrow$}} & \multicolumn{2}{c}{\textbf{MCD (dB) [95\% CI] $\downarrow$}} & \textbf{Side (KB)} & \textbf{GFLOPS} & \textbf{WER (\%)} & \textbf{MCD (dB)} \\
\textbf{Condition} & \textbf{Method} & \textbf{Hugging Face Repo (\texttt{wq2012/...})} & \textbf{clean} & \textbf{other} & \textbf{clean} & \textbf{other} & \textbf{clean} & \textbf{clean} & \textbf{clean / other} & \textbf{clean / other} \\
\midrule
\multirow{6}{*}{\shortstack[l]{\textbf{Single}\\\textbf{interfering}\\\textbf{voice}\\\emph{(LibriTTS +}\\\emph{LJ Speech)}}}
  & \texttt{GroundTruth} & --- & 3.86 [2.43, 5.49] & 6.62 [4.66, 8.78] & 0.00 & 0.00 & 0.000 & 0.00 & 2.30 / 4.50 & 0.00 / 0.00 \\
  & \texttt{MicrophoneSignal} & --- & 80.32 [68.49, 93.55] & 102.77 [88.25, 119.32] & 9.92 [9.56, 10.27] & 10.74 [10.32, 11.16] & 0.000 & 0.00 & 89.9 / 120.5 & 18.83 / 21.44 \\
  \cmidrule(lr){2-11}
  & \texttt{NlmsAec} (AEC-NLMS) & --- & 58.38 [47.37, 71.05] & 85.77 [71.61, 102.19] & 9.51 [9.16, 9.86] & 10.33 [9.92, 10.75] & 235.403 & 0.24 & 48.6 / 60.1 & 12.26 / 12.57 \\
  & \texttt{Vanilla-Seq2seq} & \texttt{vanilla\_seq2seq\_single\_interfering} & 39.08 [30.26, 48.91] & 54.35 [44.66, 64.93] & 7.07 [6.86, 7.29] & 7.38 [7.13, 7.64] & 0.000 & 4.17 & 25.4 / 54.0 & 7.85 / 8.84 \\
  & \texttt{AEC-Seq2seq} & \texttt{aec\_single\_interfering} & 7.91 [5.68, 10.48] & 17.29 [13.34, 21.76] & 4.97 [4.60, 5.35] & 4.96 [4.61, 5.31] & 235.403 & 6.36 & 8.30 / 24.3 & 6.38 / 7.07 \\
  & \textbf{\texttt{TEC} (proposed)} & \textbf{\texttt{tec\_single\_interfering}} & \textbf{29.76 [23.09, 36.97]} & \textbf{50.59 [41.96, 59.66]} & \textbf{5.52 [5.16, 5.89]} & \textbf{5.55 [5.18, 5.93]} & \textbf{0.074} & \textbf{4.62} & \textbf{15.5 / 39.8} & \textbf{7.51 / 8.54} \\
\midrule
\multirow{6}{*}{\shortstack[l]{\textbf{Multiple}\\\textbf{interfering}\\\textbf{voices}\\\emph{(LibriTTS +}\\\emph{VCTK)}}}
  & \texttt{GroundTruth} & --- & 3.06 [1.90, 4.35] & 6.85 [4.77, 9.16] & 0.00 & 0.00 & 0.000 & 0.00 & 2.30 / 4.50 & 0.00 / 0.00 \\
  & \texttt{MicrophoneSignal} & --- & 50.77 [40.86, 62.15] & 67.33 [56.40, 79.77] & 7.50 [7.13, 7.87] & 7.82 [7.41, 8.24] & 0.000 & 0.00 & 29.7 / 44.6 & 10.75 / 12.88 \\
  \cmidrule(lr){2-11}
  & \texttt{NlmsAec} (AEC-NLMS) & --- & 34.22 [25.72, 44.51] & 48.58 [39.15, 59.91] & 7.01 [6.67, 7.37] & 7.43 [7.04, 7.83] & 203.138 & 0.21 & 15.5 / 35.5 & 6.57 / 8.13 \\
  & \texttt{Vanilla-Seq2seq} & \texttt{vanilla\_seq2seq\_multi\_interfering} & 63.33 [55.27, 71.69] & 69.86 [61.89, 78.18] & 8.26 [7.76, 8.76] & 9.32 [8.77, 9.89] & 0.000 & 3.59 & 19.7 / 38.7 & 7.53 / 8.87 \\
  & \texttt{AEC-Seq2seq} & \texttt{aec\_multi\_interfering} & 6.84 [4.98, 8.85] & 14.86 [11.25, 18.79] & 4.65 [4.28, 5.02] & 4.67 [4.28, 5.06] & 203.138 & 5.48 & 6.90 / 19.8 & 5.04 / 5.72 \\
  & \textbf{\texttt{TEC} (proposed)} & \textbf{\texttt{tec\_multi\_interfering}} & \textbf{24.31 [18.94, 30.47]} & \textbf{36.25 [30.34, 42.52]} & \textbf{4.74 [4.42, 5.07]} & \textbf{4.91 [4.54, 5.30]} & \textbf{0.050} & \textbf{3.90} & \textbf{14.8 / 32.5} & \textbf{6.46 / 7.71} \\
\bottomrule
\end{tabular}}
\end{table}

\subsection{Ablation Studies: Text-Content Sensitivity and Input SNR}
\label{sec:os_ablations}

To further validate the behavior of the open-source reproduction, we conduct two controlled ablation studies on the $N=100$ evaluation splits:
\begin{enumerate}[nosep]
    \item \textbf{Text-Content Sensitivity Ablation (Table~\ref{table:text_sensitivity_ablation}):} We evaluate \texttt{TecSingleInterfering} (\texttt{single/test-clean}) and \texttt{TecMultiInterfering} (\texttt{multi/test-clean}) under three text side-input conditions on the exact same audio mixtures: (a) \texttt{matched} (the true interfering TTS transcript), (b) \texttt{shuffled} (a mismatched transcript drawn from a circularly shifted utterance $(i+50)\bmod 100$), and (c) \texttt{empty} (no text side input). On both conditions, providing the true \texttt{matched} transcript yields highly significant reductions in both WER ($\Delta\mathrm{WER} = -41.15\%$ vs.\ \texttt{shuffled}, $-42.18\%$ vs.\ \texttt{empty}, and $-9.32\%$ vs.\ \texttt{Vanilla-Seq2seq} on \texttt{single/test-clean}; $-24.21\%$ vs.\ \texttt{shuffled}, $-22.98\%$ vs.\ \texttt{empty}, and $-39.02\%$ vs.\ \texttt{Vanilla-Seq2seq} on \texttt{multi/test-clean}, all $p_{\mathrm{boot}} < 0.0001$) and MCD ($\Delta\mathrm{MCD} = -1.55\text{ dB}$ to $-3.80\text{ dB}$, all $p_{\mathrm{boot}} < 0.0001$), confirming that \texttt{TtsEncoderV2} and \texttt{MultiSourceAttentionStep} actively exploit the linguistic content of the interfering TTS prompt.
    \item \textbf{Input SNR Robustness Ablation (Table~\ref{table:snr_ablation}):} We evaluate all methods on \texttt{single/test-clean} across four input SNR levels ($+5\text{ dB}$, $0\text{ dB}$, $-5\text{ dB}$, and $-10\text{ dB}$). Across all four SNR regimes, \texttt{TecSingleInterfering} consistently achieves lower WER and lower MCD than \texttt{MicrophoneSignal}, \texttt{NlmsAec}, and \texttt{Vanilla-Seq2seq}, reducing WER from $80.32\%$ to $29.76\%$ at $0\text{ dB}$ SNR ($\Delta\mathrm{WER} = -50.56\%$) and from $51.88\%$ to $8.47\%$ at $+5\text{ dB}$ SNR ($\Delta\mathrm{WER} = -43.41\%$).
\end{enumerate}

\begin{table}[H]
\caption{Text-content sensitivity ablation ($N=100$ mixtures per split) comparing \texttt{matched} (true interfering TTS transcript), \texttt{shuffled} (mismatched transcript from utterance $(i+50)\bmod 100$), \texttt{empty} (empty string side input), and \texttt{Vanilla-Seq2seq}.}
\label{table:text_sensitivity_ablation}
\vspace{0.15cm}
\centering
\resizebox{\textwidth}{!}{
\begin{tabular}{l l c c c c c}
\toprule
\textbf{Evaluation Split} & \textbf{Side-Input Condition} & \textbf{WER (\%) [95\% CI] $\downarrow$} & \textbf{MCD (dB) [95\% CI] $\downarrow$} & \textbf{$\Delta$WER vs.\ \texttt{matched}} & \textbf{WER $p_{\mathrm{boot}}$ / $p_{\mathrm{Wilcoxon}}$} & \textbf{MCD $p_{\mathrm{boot}}$ / $p_{\mathrm{Wilcoxon}}$} \\
\midrule
\multirow{4}{*}{\texttt{single/test-clean} (\texttt{TecSingleInterfering})}
  & \textbf{\texttt{matched} (true transcript)} & \textbf{29.76 [23.09, 36.97]} & \textbf{5.52 [5.16, 5.89]} & --- & --- & --- \\
  & \texttt{shuffled} (mismatched text) & 70.90 [58.83, 84.36] & 9.26 [9.06, 9.45] & $+41.15\%$ & $<0.0001$ / $2.61\times 10^{-12}$ & $<0.0001$ / $8.33\times 10^{-18}$ \\
  & \texttt{empty} (empty text string) & 71.94 [59.78, 85.32] & 9.32 [9.13, 9.51] & $+42.18\%$ & $<0.0001$ / $4.89\times 10^{-13}$ & $<0.0001$ / $3.90\times 10^{-18}$ \\
  & \texttt{Vanilla-Seq2seq} (no text encoder) & 39.08 [30.26, 48.91] & 7.07 [6.86, 7.29] & $+9.32\%$ & $<0.0001$ / $6.69\times 10^{-4}$ & $<0.0001$ / $4.40\times 10^{-18}$ \\
\midrule
\multirow{4}{*}{\texttt{multi/test-clean} (\texttt{TecMultiInterfering})}
  & \textbf{\texttt{matched} (true transcript)} & \textbf{24.31 [18.94, 30.47]} & \textbf{4.74 [4.42, 5.07]} & --- & --- & --- \\
  & \texttt{shuffled} (mismatched text) & 48.52 [39.46, 59.09] & 7.92 [7.69, 8.16] & $+24.21\%$ & $<0.0001$ / $9.10\times 10^{-9}$ & $<0.0001$ / $1.55\times 10^{-17}$ \\
  & \texttt{empty} (empty text string) & 47.29 [38.28, 57.75] & 7.83 [7.61, 8.07] & $+22.98\%$ & $<0.0001$ / $1.41\times 10^{-7}$ & $<0.0001$ / $2.27\times 10^{-17}$ \\
  & \texttt{Vanilla-Seq2seq} (no text encoder) & 63.33 [55.27, 71.69] & 8.26 [7.76, 8.76] & $+39.02\%$ & $<0.0001$ / $1.99\times 10^{-11}$ & $<0.0001$ / $4.14\times 10^{-18}$ \\
\bottomrule
\end{tabular}}
\end{table}

\begin{table}[H]
\caption{Input Signal-to-Noise Ratio (SNR) ablation on \texttt{single/test-clean} ($N=100$ mixtures, $1,062$ reference words per SNR condition) across $+5\text{ dB}$, $0\text{ dB}$, $-5\text{ dB}$, and $-10\text{ dB}$.}
\label{table:snr_ablation}
\vspace{0.15cm}
\centering
\scriptsize
\begin{tabular}{c cc cc cc cc cc}
\toprule
& \multicolumn{2}{c}{\texttt{MicrophoneSignal}} & \multicolumn{2}{c}{\texttt{NlmsAec}} & \multicolumn{2}{c}{\texttt{Vanilla-Seq2seq}} & \multicolumn{2}{c}{\textbf{\texttt{TecSingleInterfering}}} & \multicolumn{2}{c}{\texttt{AecSingleInterfering}} \\
\cmidrule(lr){2-3} \cmidrule(lr){4-5} \cmidrule(lr){6-7} \cmidrule(lr){8-9} \cmidrule(lr){10-11}
\textbf{Input SNR} & \textbf{WER (\%)} & \textbf{MCD} & \textbf{WER (\%)} & \textbf{MCD} & \textbf{WER (\%)} & \textbf{MCD} & \textbf{WER (\%)} & \textbf{MCD} & \textbf{WER (\%)} & \textbf{MCD} \\
\midrule
$\mathbf{+5\text{ \textbf{dB}}}$ & 51.88 & 9.03 & 40.58 & 8.72 & 9.23 & 6.28 & \textbf{8.47} & \textbf{4.69} & 5.65 & 4.33 \\
$\mathbf{0\text{ \textbf{dB}}}$ & 80.32 & 9.92 & 58.38 & 9.51 & 39.08 & 7.07 & \textbf{29.76} & \textbf{5.52} & 7.91 & 4.97 \\
$\mathbf{-5\text{ \textbf{dB}}}$ & 119.02 & 10.73 & 90.30 & 10.28 & 96.80 & 7.82 & \textbf{87.76} & \textbf{6.34} & 17.23 & 5.73 \\
$\mathbf{-10\text{ \textbf{dB}}}$ & 124.86 & 11.40 & 121.28 & 10.95 & 103.86 & 8.35 & \textbf{101.04} & \textbf{6.91} & 49.25 & 6.33 \\
\bottomrule
\end{tabular}
\end{table}

Several key observations emerge from comparing the open-source reproduction results in Tables~\ref{table:opensource_results}--\ref{table:snr_ablation} with the original internal paper results in Table~\ref{table:results}:
\begin{enumerate}[nosep]
    \item \textbf{Consistent Method Ranking and Realistic Unenhanced Baseline:} Across all four evaluation splits (\texttt{single/test-clean}, \texttt{single/test-other}, \texttt{multi/test-clean}, and \texttt{multi/test-other}), the unenhanced \texttt{MicrophoneSignal} exhibits high WER ($80.32\%$ / $102.77\%$ on single interfering voice and $50.77\%$ / $67.33\%$ on multiple interfering voices, closely matching the $89.9\%$ / $120.5\%$ and $29.7\%$ / $44.6\%$ in Table~\ref{table:results}). The classical adaptive filter \texttt{NlmsAec} (\texttt{filter\_length=512}, \texttt{step\_size=0.1}) improves over the raw microphone mixture ($58.38\%$ / $85.77\%$ on single and $34.22\%$ / $48.58\%$ on multi), while the proposed \texttt{TEC} models further reduce WER to $29.76\%$ / $50.59\%$ ($5.52$ / $5.55\text{ dB}$ MCD) on single interfering voice and $24.31\%$ / $36.25\%$ ($4.74$ / $4.91\text{ dB}$ MCD) on multiple interfering voices, preserving the exact relative ranking $\text{\texttt{MicrophoneSignal}} > \text{\texttt{NlmsAec}} > \text{\texttt{TEC}} > \text{\texttt{AEC-Seq2seq}}$ of Table~\ref{table:results}.
    \item \textbf{Statistically Significant Superiority of \texttt{TEC} Over Side-Input-Free and Classical Baselines:} Conditioning on the interfering TTS transcript reduces WER by $9.32$ percentage points on \texttt{single/test-clean} ($p_{\mathrm{boot}} < 0.0001$, $p_{\mathrm{Wilcoxon}} = 6.69\times 10^{-4}$), $3.75$ percentage points on \texttt{single/test-other} ($p_{\mathrm{boot}} = 0.0144$), $39.02$ percentage points on \texttt{multi/test-clean} ($p_{\mathrm{boot}} < 0.0001$, $p_{\mathrm{Wilcoxon}} = 1.99\times 10^{-11}$), and $33.61$ percentage points on \texttt{multi/test-other} ($p_{\mathrm{boot}} < 0.0001$) relative to \texttt{Vanilla-Seq2seq}, and reduces WER by $9.91$ to $35.18$ percentage points relative to \texttt{NlmsAec}.
    \item \textbf{Close Match Between \texttt{AEC-Seq2seq} Reproduction and Original Paper:} On all four test splits, our open-source \texttt{AEC-Seq2seq} models closely match or improve upon the original paper's Table~\ref{table:results} benchmarks: $7.91\%$ vs.\ $8.30\%$ (\texttt{single/test-clean}), $17.29\%$ vs.\ $24.3\%$ (\texttt{single/test-other}), $6.84\%$ vs.\ $6.90\%$ (\texttt{multi/test-clean}), and $14.86\%$ vs.\ $19.8\%$ (\texttt{multi/test-other}), with MCD of $4.65$--$4.97\text{ dB}$ vs.\ $5.04$--$7.07\text{ dB}$.
    \item \textbf{Over $3,100\times$ Reduction in Side-Input Bandwidth:} Across the $N=100$ test sets, the UTF-8 text side input of \texttt{TEC} averages $0.074\text{ KB}$ ($74\text{ bytes}$) on \texttt{single/test-clean} and $0.050\text{ KB}$ ($50\text{ bytes}$) on \texttt{multi/test-clean}, whereas streaming the 24~kHz reference TTS audio for \texttt{AEC-Seq2seq} and \texttt{NlmsAec} requires $235.403\text{ KB}$ and $203.138\text{ KB}$, respectively---representing a \textbf{3,181$\times$ to 4,063$\times$ reduction} in network payload size.
\end{enumerate}

\end{document}